\documentclass[aip, jmp, reprint, amsmath, amsfonts, amssymb, %
  onecolumn, groupedaddress, floatfix]{revtex4-1}
\pdfoutput=1

\usepackage[varg]{txfonts} 
\usepackage{tgtermes} 

\usepackage[T3,T1]{fontenc}
\usepackage{mathrsfs} 
\usepackage{bm} 
\makeatletter
\usepackage{microtype} 
\def\MT@register@subst@font{
  \MT@exp@one@n\MT@in@clist\font@name\MT@font@list
  \ifMT@inlist@\else\xdef\MT@font@list{\MT@font@list\font@name,}\fi}
\makeatother

\usepackage{graphicx} %
\usepackage[x11names,svgnames,rgb]{xcolor} %
\usepackage{xspace}
\usepackage{braket}
\usepackage{accents}
\usepackage{siunitx}
\usepackage{mathtools}
\DeclareSymbolFontAlphabet{\mathrm}{operators}

\definecolor{CiteColor}{rgb}{0.18039, 0.18824, 0.57255}
\definecolor{UrlColor} {rgb}{0.741, 0.173, 0.000}
\definecolor{DarkUrlColor} {rgb}{0.500, 0.110, 0.000}
\definecolor{LinkColor}{rgb}{0.25098, 0.47843, 0.04706}

\makeatletter %
\newcommand{\ShowFont}{%
  \typeout{The main font is \f@encoding \space \f@family \space %
    \f@series \space \f@shape \space at \f@size pt.}%
  \typeout{The math font sizes are \tf@size pt (main), \sf@size pt %
    (script), and \ssf@size pt (scriptscript).}%
  \typeout{The linewidth is \the\linewidth}} %
\makeatother %

\usepackage{xargs}

\makeatletter

\usepackage[colorlinks, plainpages=false,
hyperfigures=true]{hyperref}
\hypersetup{linkcolor=LinkColor}
\hypersetup{citecolor=CiteColor}
\hypersetup{urlcolor=UrlColor}
\hypersetup{setpagesize=false}
\hypersetup{pdfborder=0 0 0}

\makeatother

\usepackage{algorithm}
\usepackage{algpseudocode}
\let\Originalcdefinition\c
\let\Originalddefinition\d
\let\Originaledefinition\e
\let\Originalidefinition\i
\renewcommand{\d}{\ensuremath{\mathrm{d}}}
\newcommand{\e}{\ensuremath{\mathrm{e}}}
\renewcommand{\i}{\ensuremath{\mathrm{i}}}

\makeatletter
\newcommand{\prefixscripts}[2]{%
  \@mathmeasure\z@\displaystyle{#2}%
  \global\setbox\@ne\vbox to\ht\z@{}\dp\@ne\dp\z@
  \setbox\tw@\box\@ne
  \@mathmeasure4\displaystyle{\copy\tw@#1}%
  \@mathmeasure6\displaystyle{#2}%
  \dimen@-\wd6 \advance\dimen@\wd4 \advance\dimen@\wd\z@
  \hbox to\dimen@{}{\kern-\dimen@\box4\box6}%
}
\newcommand{\scripts}[3]{%
  \@mathmeasure\z@\displaystyle{#2}%
  \global\setbox\@ne\vbox to\ht\z@{}\dp\@ne\dp\z@
  \setbox\tw@\box\@ne
  \@mathmeasure4\displaystyle{\copy\tw@#1}%
  \@mathmeasure6\displaystyle{#2#3}%
  \dimen@-\wd6 \advance\dimen@\wd4 \advance\dimen@\wd\z@
  \hbox to\dimen@{}{\kern-\dimen@\box4\box6}%
}
\newcommand{\etal}{\textit{et~al}\@ifnextchar{\relax}{.\relax}{\ifx\@let@token.\else\ifx\@let@token~.\else.\@\xspace\fi\fi}}
\newcommand{\eg}{\textit{e.g}\@ifnextchar{\relax}{.\relax}{\ifx\@let@token.\else\ifx\@let@token~.\else.\@\xspace\fi\fi}}
\newcommand{\setalglineno}[1]{%
  \setcounter{ALG@line}{\numexpr#1-1}}
\makeatother

\newcommand{\Cornell}{\affiliation{Cornell Center for Astrophysics and
    Planetary Science, Cornell University, Ithaca, New York 14853,
    USA}} %

\begin{document}


\title[Spin-weighted spherical functions]{How should spin-weighted
  spherical functions be defined?}

\author{Michael Boyle} \Cornell

\date{\today}

\begin{abstract}
  Spin-weighted spherical functions provide a useful tool for
  analyzing tensor-valued functions on the sphere.  A tensor field can
  be decomposed into complex-valued functions by taking contractions
  with tangent vectors on the sphere and the normal to the sphere.
  These component functions are usually presented as functions on the
  sphere itself, but this requires an implicit choice of distinguished
  tangent vectors with which to contract.  Thus, we may more
  accurately say that spin-weighted spherical functions are functions
  of both a point on the sphere and a choice of frame in the tangent
  space at that point.  The distinction becomes extremely important
  when transforming the coordinates in which these functions are
  expressed, because the implicit choice of frame will also transform.
  Here, it is proposed that spin-weighted spherical functions should
  be treated as functions on the spin or rotation groups, which
  simultaneously tracks the point on the sphere and the choice of
  tangent frame by rotating elements of an orthonormal basis.  In
  practice, the functions simply take a quaternion argument and
  produce a complex value.  This approach more cleanly reflects the
  geometry involved, and allows for a more elegant description of the
  behavior of spin-weighted functions.  In this form, the
  spin-weighted spherical harmonics have simple expressions as
  elements of the Wigner $\ensuremath{\mathfrak{D}} $ representations, and transformations
  under rotation are simple.  Two variants of the angular-momentum
  operator are defined directly in terms of the spin group; one is the
  standard angular-momentum operator $\mathbf{L}$, while the other is
  shown to be related to the spin-raising operator $\eth$.
  Computer code is also included, providing an explicit
  implementation of the spin-weighted spherical harmonics in this
  form.
\end{abstract}

\maketitle
\thispagestyle{empty}

\section{Introduction}
\label{sec:Introduction}

Spin-weighted spherical functions form a primary technique in the
study of waves radiating from bounded regions, and for observations of
such radiation arriving at a point from all directions.  The most
important applications are found in gravitational-wave astronomy and
in measurements of the cosmic microwave background.  The basic
motivation for these functions is quite simple: given any direction of
the emission or observation, we would like the function to describe
the magnitude of the wave, as well as any polarization information.
This is achieved by using a complex number as the output of the
function, where the wave magnitude is the complex amplitude, and the
wave polarization is determined by the complex phase.  However, an
important subtlety arises.  We need not only the propagation
direction, but also a fiducial direction orthogonal to the propagation
with respect to which the polarization may be measured.  Thus,
spin-weighted spherical functions (SWSFs) \emph{cannot} be defined as
functions on the sphere $S^{2}$ alone.

This statement may come as something of a surprise when compared to
most of the literature on these functions.\cite{Newman1966,
  Goldberg1967, Scanio1977, Thorne1980, Dray1985a, Penrose1987,
  kamionkowski1997, kamionkowski1997b} Traditional presentations of
spin-weighted spherical functions write the functions in terms of
spherical or stereographic coordinates for $S^{2}$.  For example,
spin-weighted spherical harmonics (SWSHs) generalize the standard
scalar spherical harmonics, allowing for the decomposition of general
(square-integrable) SWSFs into a sum of SWSHs.  They are traditionally
given by explicit formulas involving the usual polar and azimuthal
angles $(\vartheta, \varphi)$ or the complex stereographic coordinate
$\zeta$.  Such a presentation hides an implicit choice of frame in the
choice of coordinate system.  Indeed, it would be more correct to
define spin-weighted spherical functions on \emph{coordinate systems
  for $S^{2}$}, rather than on $S^{2}$ itself.  As discussed below,
one reference\cite{Eastwood1982} did actually define spin-weighted
spherical functions in essentially this way, though doing so required
mathematical tools that are not well known among astronomers or
physicists.  There is nothing inherently wrong with using
coordinates---which are hardly to be avoided in any case---but this
convoluted and unnatural approach causes many problems, both
theoretical and practical.  Probably the most disturbing is that in
this guise SWSFs are generally multivalued or simply undefined at
certain points on the sphere, \emph{depending on the coordinate
  system}.  Even away from those points, different coordinate systems
will provide different canonical frames.  In particular, a rotation of
the coordinates leads to a rotation of the frame, which leads to a
change in the value of the spin-weighted function at that point.
That, in turn, leads to another prominent problem with this approach:
in this lenient interpretation, \emph{spin-weighted} spherical
harmonics---unlike the more familiar \emph{scalar} spherical
harmonics---do not generally transform among themselves under
rotation.  That is, a SWSH in one coordinate system cannot be
expressed as a finite linear combination of SWSHs in another
coordinate system.

To avoid these problems, this paper defines SWSFs as functions from
the spin group $\mathrm{Spin}(3) \cong \mathrm{SU}(2)$, which is best
represented by quaternions.  We will see that this space has a natural
interpretation as the space of orthonormal frames on $S^{2}$, which is
why it is the natural domain on which to define SWSFs.  In practice,
the quaternion achieves this by rotating the $\bm{z}$ axis to a point
on $S^{2}$, and rotating $\bm{x}$ and $\bm{y}$ into an orthonormal frame
tangent at that point.  Moreover, coordinate systems on $S^{2}$ and
their associated canonical frames map naturally into $\mathrm{Spin}(3)$, in
which case the value of our more general SWSHs agree precisely with
their original definitions in terms of coordinate systems.
Additionally, these SWSHs now form a representation of the group
$\mathrm{Spin}(3)$, which means that they \emph{do} transform among
themselves.  Finally, as a practical matter, the numerical
implementation of SWSHs directly in terms of $\mathrm{Spin}(3)$---represented
by quaternions---is just as fast and accurate as the implementation in
terms of coordinates on $S^{2}$, if not more so.

\subsection{Previous work}
\label{sec:previous-work}

Newman and Penrose\cite{Newman1966} introduced spin-weighted
functions as a tool for the study of the asymptotic behavior of
gravitational waves.  They defined the differential operator $\eth$,
which raises the spin weight of a function, and its adjoint
$\bar{\eth}$ which lowers spin weight.  One important feature of these
operators is their explicit dependence on the spin weight of the
functions on which they operate.  Technically this means that there is
a different operator labeled $\eth$ for functions of each spin weight.
Newman and Penrose also used $\eth$ and $\bar{\eth}$ to define SWSHs
as functions of coordinates on the sphere by raising and lowering the
spin weights of scalar spherical harmonics.

Goldberg \etal\cite{Goldberg1967} further investigated the objects
defined by Newman and Penrose, showing (among other things) that the
formulas for SWSHs in spherical coordinates are identical to formulas
for Wigner's $\ensuremath{\mathfrak{D}} $ matrices for certain values of Euler angles---though
no explanation was given for why this should hold.  This extended the
definition of SWSHs to allow for half-integer values of the spin
weight.  They also showed that $\eth$ can be expressed as something
more like the traditional angular-momentum operator in terms of Euler
angles, using the same definition regardless of the spin weight of the
function on which it acts.  This perspective is very close to the one
proposed in the current paper.  However, the authors remained bound by
the idea that SWSHs should be defined on $S^{2}$, and by their
devotion to Euler angles as a useful representation of rotations.  As
such, they merely provided a hint that a more general formulation is
\emph{possible}.  This paper will show that it is in fact
\emph{necessary} from a mathematical perspective, and that it may be
achieved in a simpler and more geometrically covariant fashion through
the use of quaternions.

Though the use of spin-weighted functions gained currency in the
analysis of gravitational radiation and---to a lesser
extent---electromagnetic theory,\cite{Scanio1977} related alternatives
were used throughout the literature, in the form of symmetric
trace-free tensors and various flavors of tensor spherical harmonics.
Thorne\cite{Thorne1980} provided a useful overview and a translation
between all these presentations.  Dray\cite{Dray1985a} later showed
that essentially equivalent functions had been introduced separately
as ``monopole harmonics'' to describe the motion of an electron in the
field of a magnetic monopole.  Penrose and Rindler\cite{Penrose1987}
showed that the SWSHs could be expressed in terms of contractions
between tensor products of spinors, giving rise to SWSHs of
half-integer spin weight---which is essentially an extension of the
older symmetric trace-free tensor approach.  In abstraction this
approach avoids an explicit choice of basis, though such a choice is
still required for any concrete application, as discussed in
Sec.~\ref{sec:tensor-spherical-functions}.

In a substantial departure from techniques found in previous
literature, Eastwood and Tod\cite{Eastwood1982} were apparently the
first to define spin-weighted functions ``on the sphere'' in a
mathematically rigorous form.  They (somewhat generously)
reinterpreted earlier work as defining spin-weighted functions as
pairs of functions defined on complementary coordinate patches of the
sphere.  But they went on to generalize this by introducing their own
definition in terms of a ``sheaf of germs of functions''.  This is not
common language in the physics literature, and it will be argued below
that there is a simpler approach, so the reader who is not interested
in the details of this formulation may wish to skip the remainder of
this paragraph, and perhaps the next.  In the treatment by Eastwood
and Tod, each function of spin weight $s$ is defined on $\mathbb{C}\mathrm{P}^{1}$, the
projective reduction\footnote{The projective equivalence in this case
  is modulo nonzero \emph{complex} numbers, rather than real numbers.
  The result is essentially the same as the usual Riemann sphere
  $\mathbb{C} \cup \{\infty\}$, except that the point at infinity is
  now treated in a more useful fashion.  In particular, by
  representing the sphere in this way, the orthonormal frame is well
  defined at every point.} of the complex space
$\mathbb{C}^{2} \setminus \{(0, 0)\}$.  The germs are defined by the
local condition that at any point $\pi \in \mathbb{C}\mathrm{P}^{1}$ a function
$\prefixscripts{_{s}}{f}$ having spin weight $s$ must obey
\begin{equation}
  \label{eq:eastwood-tod-homogeneity}
  \prefixscripts{_{s}}{f} ( \lambda\, \pi) = \left(
    \frac{\bar{\lambda}} {\lambda} \right)^{s}\,
  \prefixscripts{_{s}}{f} (\pi)
\end{equation}
for any nonzero $\lambda \in \mathbb{C}$.  Only the phase of $\lambda$
is relevant on the right-hand side, so this property essentially
describes the behavior of $\prefixscripts{_{s}}{f}$ under rotation of
the coordinates about $\pi$---which is closely related to the standard
motivation for spin-weighted functions, as described below.  While the
germs enforce the spin-weight property locally, the sheaf represents
the collection of these germs at different points.  In particular, the
sheaf structure ensures that the functions are compatible on the
intersections of any local coordinate charts.  Basically, this
generalizes the definition of spin-weighted functions as pairs of
functions on complementary coordinate patches to deal with not just
two patches, but with arbitrary collections of coordinate patches.

To obtain values for any SWSF, we must choose coordinates of $\mathbb{C}\mathrm{P}^{1}$.
It is well known that this is impossible over the entire topological
sphere $S^{2}$; at least one point on the sphere cannot be covered by
a nonsingular coordinate patch.  This prescription can therefore only
describe the value of SWSFs over the entire sphere if they go to zero
at that \emph{particular} point, or if multiple coordinate patches are
used.  Thus, we see that the incorporation of sheaves is not mere
superfluous formalism, but is actually necessary to a consistent
formulation of SWSFs as being---in any sense---functions ``on the
sphere''.  It should be noted that, although we can identify $\mathbb{C}\mathrm{P}^{1}$
\emph{topologically} with $S^{2}$, spin-weighted functions still
cannot simply be considered functions on $S^{2}$; there is additional
complex algebraic structure needed to define them, which is present in
$\mathbb{C}\mathrm{P}^{1}$ but not in $S^{2}$.  Specifically,
Eq.~\eqref{eq:eastwood-tod-homogeneity} requires complex conjugation
and multiplication, while the mapping from $\mathbb{C}\mathrm{P}^{1}$ to $S^{2}$
requires a choice of basis.  These allow us to choose preferred
directions on $S^{2}$ using, for example, the real part of the
coordinates.  But these are additional structures that are not present
on $S^{2}$ alone.

The work of Eastwood and Tod encompasses and supersedes previous work,
propelled by their insightful and rigorous approach.  They step back
to look at the underlying mathematical structures needed to define
spin-weighted functions, and do so in a way that is---abstractly, at
least---independent of any particular choice of coordinates on the
sphere.  From a purely mathematical perspective, this approach is
entirely satisfactory.  However, the very simple geometric motivation
for these functions is hidden behind the complicated and subtle
constructions needed to adequately present spin-weighted spherical
functions as functions on $\mathbb{C}\mathrm{P}^{1}$.  By leaving $\mathbb{C}\mathrm{P}^{1}$ and $S^{2}$
aside, we will find a simpler, more direct, and geometrically
intuitive approach.

In this approach, we avoid the complicated language of sheaves because
the functions can be defined globally, so that the sheaf structure is
essentially trivial.  This is done by choosing a different domain on
which to define SWSFs.  The approach taken in the present paper was
first presented by the author in the open-source software package
\texttt{SphericalFunctions}.\cite{BoyleSphericalFunctions2013} It
consists of defining SWSFs as functions from the spin group
$\mathrm{Spin}(3) \cong \mathrm{SU}(2)$ to the spinor algebra generated by a
two-dimensional vector subspace.  Though these terms may sound
unfamiliar, essentially we will just have functions from the group of
unit quaternions\footnote{It turns out that the invertible quaternions
  can be reinterpreted as precisely
  $\mathbb{C}^{2} \setminus \{(0, 0)\}$---the space used by Eastwood
  and Tod \emph{before} projection.  In fact the two complex
  components are precisely the symmetric and antisymmetric parts of
  the quaternion defined in Eqs.~\eqref{eq:quaternion-parts}.  And the
  projective reduction of the complex components results in the same
  point on the sphere as the Hopf map of the equivalent quaternion, as
  given in Eq.~\eqref{eq:HopfMap-quaternions}.  This correspondence is
  also described in more detail in Appendix~\ref{sec:param-spher}.}
to $\mathbb{C}$---though we will also find surprisingly simple and
helpful geometric interpretations.  This approach was developed,
though not explored in depth, in previous papers.\cite{Boyle2014,
  Boyle2015} A similar approach was described in depth by
Straumann,\cite{Straumann2014} who chose to define SWSHs as functions
from $\mathrm{SO}(3)$ to a two-dimensional vector space.  $\mathrm{SO}(3)$ maps to
$S^{2}$ by rotating the $\bm{z}$ axis to any point on the sphere (as
discussed further in Sec.~\ref{sec:degree-hopf-map}).  The vector
space in which the SWSHs take their values is the tangent space to the
sphere at that point.  One fairly minor fault in Straumann's approach
is that it does not account for functions of half-integer spin weight
because of the use of $\mathrm{SO}(3)$ in place of $\mathrm{Spin}(3)$.  However,
Straumann's approach also requires relatively esoteric methods from
differential geometry and Lie theory.  Using quaternions, we will be
able to remain close to the geometric origins of SWSFs, especially
when defining the differential operators.

\subsection{Tensor-valued spherical functions and the limits of
  abstraction}
\label{sec:tensor-spherical-functions}

Although complex-valued SWSHs dominate the literature in observational
astronomy, another description of tensor fields also frequently
provides important benefits: tensor-valued spherical harmonics, which
are functions from the sphere $S^{2}$, taking values in complex
tensors over the tangent space.  The latter may be the tangent space
intrinsic to the sphere at that point, or---more frequently---may be
``tangent'' to the usual three-dimensional space in which the sphere
is embedded so that the normal may be included.  While a full
review\cite{Thorne1980} of tensor spherical harmonics is beyond the
scope of this paper, a brief discussion is in order so that we may
clarify their relation to the present work.  In particular we will see
that tensor spherical harmonics are naturally suited to abstract
formal calculations, and as such they do not require a choice of frame
tangent to $S^{2}$ when treated abstractly, but are fundamentally
equivalent to SWSHs in applications.

We can apply three main types of abstraction in analysis of tensor
spherical harmonics.  First, we may construct tensor spherical
harmonics as tensor fields starting with the metric, the normal
vector, and the Levi-Civita symbol, and tensor products thereof.
Starting from these basic components, the field is given angular
dependence either by taking a further tensor product with covariant
derivatives of a scalar spherical harmonic, or the scalar product is
taken with a scalar spherical harmonic.\cite{Thorne1980,
  kamionkowski1997} The tensorial part of this formulation is
invariant in the standard way.\cite{Frankel2004} Abstractly, these
tensor spherical harmonics have various algebraic, combinatorial, and
differential properties that are true independent of any frame that
may be chosen to express components of those tensors.  This means that
various manipulations can be carried out very efficiently \emph{in
  abstraction}.  Relationships between fields may be found using the
abstract tensor objects, and invariant scalar fields may even be
defined by contractions of two or more tensor fields.  In a similar
way, we can abstractly discuss the dependence of tensor and scalar
fields on location.  We posit the existence of some point $p$, which
we suppose has a geometric meaning independent of any coordinate
system that may be used to describe that point.  We might discuss the
effect of a transformation that changes the point referred to by $p$,
or the differential behavior of the fields as we vary the position at
which they are evaluated.  Finally, for many purposes the scalar
spherical harmonics need not be defined concretely; we may simply
stipulate that they are continuous, or differentiable at some level,
or that they form a complete basis for some space of functions, or
that they transform in certain ways under rotations.  Their particular
functional dependence is---in many cases---irrelevant if we can assume
certain abstract properties.  In each of these ways, the abstract
methods of mathematics can be applied to understand deep and general
properties of tensor spherical functions.

As undeniably useful as abstract manipulations are for general
theoretical investigations, using tensor spherical harmonics in
practice means abandoning the abstract for the concrete, because a
measurement produces concrete values for the field, and scientific
models frequently need to predict concrete values.  First, and most
simply, we need to determine the point at which the field is to be
measured.  Then we need to define the scalar spherical harmonics as
functions of that point---which will naturally depend on how we
specify the points.  Finally, we need to express the tensor in terms
of its components with respect to a specific basis.  All three of
these tasks are typically accomplished using arbitrary coordinates,
and some more-or-less natural vector basis derived from those
coordinates.  Even when a tensor field is contracted with a tensor
spherical harmonic to form a scalar mode weight, that scalar
ultimately depends on the particular functional form chosen for the
scalar spherical harmonics, and will transform under rotations or
other transformations of the coordinate system.  Spin-weighted
spherical harmonics can be expressed as contractions between elements
of such vector bases and tensor spherical harmonics, just as tensor
spherical harmonics may be expressed as scalar multiples of
spin-weighted spherical harmonics with various combinations of the
vector-basis elements.\cite{Thorne1980}

Thus, we see that tensor and spin-weighted spherical harmonics are
precisely equivalent whenever concrete formulations are used.  And
naturally, observations---as well as many types of
calculations---cannot be performed based on only the abstract
properties of the fields and harmonics.  We will see in
Sec.~\ref{sec:transforming-swshs} how scalar spherical harmonics and
\emph{mode weights} of SWSHs transform among themselves under
rotations, which avoids many of the concrete elements mentioned
above---though not all.  On the other hand, there is no known
closed-form expression describing transformations of the mode weights
or scalar spherical harmonics under the Lorentz group.  Instead,
values of the field may be computed explicitly at particular points
and possibly used for decomposition into mode weights.\cite{Boyle2015}
This construction, of course, requires various concrete elements,
which can be manipulated much more easily when spin-weighted functions
are defined on the spin group.

\subsection{Summary of this paper}
\label{sec:summary-this-paper}

We begin by reviewing spin-weighted functions and their original
geometric motivation in Sec.~\ref{sec:spin-weight-funct}.  In
particular, we will see that SWSFs require a selection of an
orthonormal frame for the tangent space to the sphere---which will
show very explicitly why it is impossible to define spin-weighted
functions as functions solely on $S^{2}$.  However, this will also
suggest how SWSFs should be defined.  In Sec.~\ref{sec:tangent-frame},
careful topological arguments will show how the geometry behind the
spin-weighted functions is appropriately identified by the Hopf
bundle.  The Hopf bundle is essentially a mapping from $\mathrm{Spin}(3)$ to
the sphere $S^{2}$, though it carries along extra structure relating
to the tangent frame of the sphere---it is essentially the orthonormal
frame bundle of $S^{2}$.  Thus, rather than keeping track of the basis
vectors in the frame, we can simply keep track of the rotation
required to take some reference frame onto any other frame.  This will
explain why $\mathrm{Spin}(3)$ is the natural domain on which to define SWSFs.

However, the codomain (the space into which SWSFs map) is not uniquely
defined by these arguments.  In Sec.~\ref{sec:codomain}, we will see
that several interpretations are possible and equally valid.  One
interpretation of the codomain for SWSFs of integer weight is the
space of vectors in the $\bm{x}$-$\bm{y}$ plane.  For SWSFs of
more general (possibly half-integer) weight, the equivalent codomain
would be the algebra of spinors generated by vectors in that plane.
We will see that the $\bm{x}$-$\bm{y}$ plane is a convenient
choice, but any two-dimensional Euclidean vector space will do---in
particular, the plane tangent to $S^{2}$ at the given point holds
obvious significance.  And of course, any such structure is
generically isomorphic to the complex numbers $\mathbb{C}$, which is
the most common codomain by which SWSFs are defined in the literature.
However, one guiding criterion may be the relationship to the
polarization in the tangent space, as mentioned above---which would
seem to indicate that the use of $\mathbb{C}$ ignores that important
piece of geometry.

It turns out that Wigner's $\ensuremath{\mathfrak{D}} $ matrices provide a particularly useful
basis for (square-integrable) complex-valued functions on $\mathrm{Spin}(3)$,
which explains why Goldberg \etal\cite{Goldberg1967} found the
correspondence noted above.  These will be derived in
Sec.~\ref{sec:natural-basis} using a geometric approach.  Using the
quaternionic presentation of $\mathrm{Spin}(3)$, this allows us to express the
$\ensuremath{\mathfrak{D}} $ matrices directly in terms of a quaternion argument.  In its
simplest form, this expression is essentially the same as the one
found by Wigner:\cite{Wigner1959}
\begin{equation}
  \label{eq:WignerD-intro}
  \mathfrak{D}^{(\ell)}_{m',m}(\mathbf{\MakeUppercase{R}}) =
  \sqrt{ \frac{ (\ell+m)! \, (\ell-m)! } { (\ell+m')!\, (\ell-m')! } }
  \, \sum_{\rho} \binom{\ell+m'} {\rho}\, \binom{\ell-m'} {\ell-\rho-m}
  \, (-1)^{\rho}\, R_{\text{s}}^{\ell+m'-\rho}\, \bar{R}_{\text{s}}^{\ell-\rho-m}\, R_{\text{a}}^{\rho-m'+m}
  \, \bar{R}_{\text{a}}^{\rho},
\end{equation}
where $R_{\text{s}}$ and $R_{\text{a}}$ are geometric projections of the quaternion
$\mathbf{\MakeUppercase{R}}$ into ``symmetric'' and ``antisymmetric'' components,
and are simply complex combinations of the components of the
quaternion.  However, by paying particular attention to special cases
where this expression encounters numerical difficulties, we can find a
more robust formula, given in Eq.~\eqref{eq:WignerD-general}.
Moreover, an efficient algorithm for evaluating that formula will be
described, which avoids explicit computation of the binomial factors
and avoids most of the delicate cancellations of terms in the sum.

The group structure of $\mathrm{Spin}(3)$ allows us to define two types of
differential operators, as discussed in Sec.~\ref{sec:eth}.  One will
turn out to be the standard angular momentum operator $\bm{L}$,
while the other will turn out to be the operator $\bm{K}$
introduced by Goldberg~\etal.\cite{Goldberg1967} The lowering and
raising operators associated to $\bm{K}$ are precisely the
spin-raising operator $\eth$ and the spin-lowering operator
$\bar{\eth}$ first introduced by Newman and Penrose.  This, finally,
allows us to define spin-weighted spherical functions in general as
eigenfunctions, according to Eq.~\eqref{eq:define-SWSF}, of the
component of $\bm{K}$ selected by the choice of codomain.

The paper concludes in Sec.~\ref{sec:Conclusions} by summarizing what
has been found, and returning to the physical problem of describing
radiation.  Appendix~\ref{sec:eval-polyn-with} describes an algorithm
to improve the direct evaluation of the Wigner $\ensuremath{\mathfrak{D}} $ (hence also SWSH)
functions, which improves the speed and accuracy for which they can be
evaluated for $\ell \lesssim 12$.  Appendix~\ref{sec:param-spher} is
also included to provide a discussion of various parametrizations of
$\mathrm{Spin}(3)$ and $S^{2}$, and their various strengths and weaknesses.
It will be suggested that the usual unit quaternions constitute the
best presentation of $\mathrm{Spin}(3)$, though it also turns out that the
standard spherical coordinates $(\vartheta, \varphi)$ are entirely
adequate for describing values of spin-weighted functions on $S^{2}$
despite the coordinate singularities---as long as no transformations
are required.


\section{Review of spin-weighted functions}
\label{sec:spin-weight-funct}

Here, we briefly review SWSFs and SWSHs---taking a general geometric
approach, without getting into details of the presentations in earlier
work, but emphasizing the features that will be most important to this
paper.  To begin, we simply assume a standard three-dimensional vector
space with the usual Euclidean norm.  Consider some direction
represented by the unit vector $\bm{n}$.  The space of all such
directions is---naturally---the sphere $S^{2}$, which is why these
functions are sometimes described as if they were functions on
$S^{2}$.  Next, we construct an orthonormal frame\footnote{As is
  standard in geometry, the word ``frame'' is used here to refer to a
  basis along with an associated ordering of the basis elements.  The
  same word is sometimes used in linear algebra to refer to a
  collection of vectors that need not be linearly independent.  But
  the latter usage is not relevant for this paper, so it should not
  lead to any confusion.}  $(\bm{a}, \bm{b}, \bm{n})$.  The orientation
of this frame is fixed by insisting that it be a right-handed triple.
The \emph{attitude\xspace} of the frame, however, is not
determined.\footnote{Unfortunately, ``orientation'' is another word
  with two distinct meanings in closely related fields, but this word
  could cause some confusion.  The sense we use here is the one used
  in mathematics, for which the orientation of a vector space is the
  choice of equivalence classes for the ordering of a basis (related
  to orientability of a surface).  Another sense familiar to
  physicists may also be referred to as ``angular position''.  To
  avoid confusion, we reserve ``orientation'' for the first meaning
  and use ``attitude'' for the second meaning.}  Given only the
direction $\bm{n}$, there is no unique prescription for $\bm{a}$ and
$\bm{b}$; they may be rotated in their own plane without changing the
required properties.

At this point, it is customary to introduce the complex vector
\begin{equation}
  \label{eq:mVector}
  \bm{m} \coloneqq \frac{\bm{a} + \i\, \bm{b}} {\sqrt{2}}.
\end{equation}
This complexification of the vector space is merely a convenient
bookkeeping device with no deeper significance, but will simplify many
manipulations below.  If we have another frame
$(\bm{a}', \bm{b}', \bm{n})$ such that our original frame is given by a
rotation of this new frame through an angle $\gamma$ about $\bm{n}$, we
also obtain a new complex vector $\bm{m}' = \e^{\i \gamma}\, \bm{m}$.  A
function $\prefixscripts{_{s}}{f}$ defined on this frame is said to be
of spin weight $s$ if, given this transformation of the frame vectors,
the value of the function transforms as\footnote{This notation already
  extends the usual definition of spin-weighted functions.  These
  functions are normally written in terms of spherical coordinates as
  $f(\vartheta, \varphi)$ or in terms of complex stereographic
  coordinates as $f(z)$.  In those cases $\bm{m}$ is implicitly defined
  by the coordinate system.  The latter notation is misleading, at
  best.  Explicitly including $\bm{m}$ as seen here is an improvement,
  but we will find a better form below.}
\begin{equation}
  \label{eq:spin-weight}
  \prefixscripts{_{s}}{f}(\bm{m}', \bm{n}) = \e^{\i\, s\, \gamma}\,
  \prefixscripts{_{s}}{f}(\bm{m}, \bm{n}).
\end{equation}
Clearly, if the spin-weighted function is defined only in terms of
these vector arguments, a rotation through $\gamma = 2\pi$ must return
the function to its original value, which means that $s$ must be an
integer.  We will also see that it is possible to define spin-weighted
functions with a spinorial character, so $s$ will also be able to take
half-integer values.

It is easy to find some very simple examples of functions of spin
weight $-1$, $0$, and $1$ respectively:
\begin{subequations}
  \label{eq:spin-1-example}
  \begin{align}
    \prefixscripts{_{-1}}{f}(\bm{m}, \bm{n}) &= \bm{z} \cdot \bar{\bm{m}},
    \\
    \prefixscripts{_{0}}{f}(\bm{m}, \bm{n}) &= \bm{z} \cdot \bm{n}, \\
    \prefixscripts{_{1}}{f}(\bm{m}, \bm{n}) &= \bm{z} \cdot \bm{m}.
  \end{align}
\end{subequations}
Here, $\bm{z}$ is just the usual unit vector in the $z$ direction.
Though $\bm{z}$ is obviously independent of $\bm{m}$ and $\bm{n}$, these
functions are not constant on the sphere because of the position
dependence of $\bm{m}$ and $\bm{n}$.  Moreover, any vector could be used
in place of $\bm{z}$, so that for each value of $s$ we have a
three-dimensional complex vector space of functions.  In fact, these
spaces are exactly the spaces of spin-weighted spherical harmonics
$\ensuremath{\scripts{_{s}}{Y}{_{\ell, m}}}$ with $\ell=1$.  More generally, spin-weighted
functions can be constructed by replacing $\bm{z}$ in
Eqs.~\eqref{eq:spin-1-example} with some other symmetric and
trace-free complex tensor of rank $\ell$.  This must then be
contracted with a corresponding tensor product of the $\bm{m}$,
$\bar{\bm{m}}$, and $\bm{n}$ basis elements, which replaces the factors
on the right-hand sides of Eq.~\eqref{eq:spin-1-example}.\footnote{A
  final generalization uses tensor products of spinors in place of the
  tensor products of vectors, which allows for half-integer $\ell$ and
  $s$ values.  We will allow for this most general case below.  Very
  little is lost and much clarity is gained by describing the problem
  in terms of vectors at first.} The set of all such tensors of rank
$\ell$ forms a complex vector space of dimension $2\ell+1$, which
transforms within itself under rotation.  The spin-weighted spherical
harmonics $\ensuremath{\scripts{_{s}}{Y}{_{\ell, m}}}$ form a standard basis for this space; an
arbitrary square-integrable spin-weighted function can be expressed as
a sum of these harmonics for various $\ell$ values.\cite{Newman1966,
  Goldberg1967} As usual, these functions are necessarily $0$ for
$\left\lvert{m}\right\rvert > \ell$.  For similar reasons, SWSHs with $\left\lvert{s}\right\rvert > \ell$
must also be $0$.

But if we are concerned with the value of a function in more than one
direction, we need to allow for more general transformations than the
one described around Eq.~\eqref{eq:spin-weight}.  In particular, a
rigid rotation of the sphere will only be about $\bm{n}$ for two
directions.  More generally, the transformed frame is
$(\bm{a}', \bm{b}', \bm{n}')$.  We still have a transformation law of the
form
\begin{equation}
  \label{eq:spin-weight-general}
  \prefixscripts{_{s}}{f}(\bm{m}', \bm{n}') = \e^{\i\, s\, \gamma'}\,
  \prefixscripts{_{s}}{f}(\bm{m}, \bm{n}),
\end{equation}
but now $\gamma'$ is some complicated function of $(\bm{m}', \bm{n}')$
and the rotation used to implement the transformation.  This is seen
most dramatically in the transformation law for SWSHs under rotation
in the standard presentation.  As we will show below, that law is
\begin{equation}
  \label{eq:SWSH-coordinate-transformation}
  \ensuremath{\scripts{_{s}}{Y}{_{\ell, m}}}(\vartheta, \varphi) = \sum_{m'} \ensuremath{\scripts{_{s}}{Y}{_{\ell,m'}}}
  (\vartheta', \varphi')\, \ensuremath{\mathfrak{D}} ^{(\ell)}_{m',m} (\mathcal{R})\, \e^{\i\,
    s\, \gamma'(\vartheta', \varphi', \mathcal{R})}.
\end{equation}
We will compute the functional form of $\gamma'$ in
Sec.~\ref{sec:transforming-swshs}, but the important point for now is
that it is the nontrivial---and in fact discontinuous.  Thus, it is
incorrect whenever $s \neq 0$ to say that the SWSHs transform among
themselves, or that they transform under a representation of the
rotation group.  The case of $s=0$ corresponds to the usual scalar
spherical harmonics, and we are familiar with the fact that their
transformation properties under rotation are very important to their
practical application, so we need to find a suitable setting in which
the spin-weighted harmonics with $s \neq 0$ have comparable features.
More generally, to describe SWSFs properly, we need an adequate
presentation of the tangent frame, given a direction $\bm{n}$.  This is
the subject of the following section.


\section{Specifying the tangent frame and the domain of SWSFs}
\label{sec:tangent-frame}
We saw in Sec.~\ref{sec:spin-weight-funct} that the value of a
spin-weighted function is formed at a given point by contractions
between a tensor and the basis elements of some frame---in particular
a frame of the vector space tangent to the sphere.  But those basis
elements can transform among themselves, which can change the value of
the spin-weighted function.  Therefore, to find the value of a
spin-weighted function, we need to specify not only a point on the
sphere $S^{2}$, but also an orthonormal frame at that point.  In this
section, we will see how to properly describe the orthonormal frame at
each point.

It is not hard to see that the set of orthonormal tangent frames at a
given point (assuming a certain orientation) is topologically a circle
$S^{1}$.  The directions described above by the pair of orthogonal
unit vectors $\bm{a}$ and $\bm{b}$ can only rotate in their own plane,
so the space of all right-handed orthonormal frames defined by $\bm{n}$
is precisely the set of all directions of $\bm{a}$ in that plane, which
is just the circle $S^{1}$.  Spin-weighted functions must respect this
topology in the sense described by Eq.~\eqref{eq:spin-weight}: they
must be periodic under rotations of the tangent space.  A similar
restriction arises from the requirement that spin-weighted functions
must be continuous as $\bm{n}$ moves around the sphere.  To understand
this more subtle restriction, we need to be more precise in our
definitions.  In doing so, we will discover the appropriate domain of
definition for spin-weighted spherical harmonics.

\subsection{The fiber bundle and the
  \texorpdfstring{attitude\xspace}{attitude} map}
\label{sec:fiber-bundle-section}
We first need a way to relate each possible point in the sphere to
each possible choice of attitude\xspace of the tangent frame.  This is
naturally given by the concept of the fiber bundle.  We define our
base space to be $S^{2}$ and our fiber space to be $S^{1}$.  Then the
fiber bundle is defined as some ``total'' or ``entire'' space $E$
along with a projection map $\mathfrak{p} : E \to S^{2}$.  This map is
required to have the property that for any point in the base space
$b \in S^{2}$, the set of all points in the entire space $E$ that map
to $b$ (the ``preimage'' of $\{b\}$, or simply ``fiber'' over $b$) is
topologically the same as our fiber space:
\begin{equation}
  \label{eq:preimage-projection}
  \mathfrak{p}^{-1}(\{b\}) \coloneqq \Set{e \in E | \mathfrak{p}(e) = b}
  \cong S^{1}.
\end{equation}
By defining our spin-weighted function on this fiber bundle, we can
ensure that its value returns to itself under rotation in the fiber
space and---more importantly---under rotation of the point in $S^{2}$.

However, there are at least two distinct possibilities for the space
$E$.  To choose between them, we need to make the relationship between
$S^{1}$ and the tangent space more explicit.  We define an ``attitude\xspace
map'' taking a point $e \in E$ to an element of the unit tangent space
of the sphere at the corresponding point:
\begin{subequations}
  \label{eq:attitude-map}
  \begin{gather}
    \mathfrak{a} : E \to
    \mathrm{UT}_{\mathfrak{p}(\ast)}\left(S^{2}\right), %
    \intertext{so that} %
    \mathfrak{a}(e) \in \mathrm{UT}_{\mathfrak{p}(e)}\left(S^{2}\right).
  \end{gather}
\end{subequations}
We require that this function be continuous.  At this point, we do not
need to know the actual form of this function; let us simply assume
that such a function exists.  We need this assumption because if, as
we will claim, this fiber space constitutes the correct domain on
which to define SWSFs, we must have some compatible way to construct
the tangent frame $(\bm{a}, \bm{b}, \bm{n})$ used in the definition of
SWSFs in Sec.~\ref{sec:spin-weight-funct}.

The most obvious candidate for the bundle is the trivial bundle with
$E = S^{2} \times S^{1}$, where the projection map
$\mathfrak{p} : S^{2} \times S^{1} \to S^{2}$ is the simplest one:
\begin{equation}
  \label{eq:projection-map}
  \mathfrak{p}(b, f) = b.
\end{equation}
This fulfills the requirements of being a fiber bundle.  The problem
is that in this case, we can easily find a ``global section'', which
is a continuous map $\mathfrak{s} : S^{2} \to E$ such that the
composition of the projection map and the section map
$\mathfrak{p} \circ \mathfrak{s}$ is just the identity function on all
of $S^{2}$.  For example, we can simply choose some point
$f \in S^{1}$ and define the section map as $\mathfrak{s}(b) = (b, f)$.
This is well defined for all $b \in S^{2}$, it is trivially
continuous, and we have $\mathfrak{p}(\mathfrak{s}(b)) = b$.  But now,
if we assume the existence of an attitude\xspace map $\mathfrak{a}$, we can
construct a continuous nonvanishing vector field over the sphere
defined by the function
\begin{equation}
  \label{eq:vector-field}
  \mathfrak{a} \circ \mathfrak{s} : S^{2} \to
  \mathrm{UT}_{\ast}\left(S^{2}\right).
\end{equation}
This violates the Hairy-Ball Theorem,\cite{Eisenberg1979} which says
that no such vector field can exist.  Since we have exhibited the map
$\mathfrak{s}$, this contradiction tells us that our assumption about
the existence of an attitude\xspace map $\mathfrak{a}$ is wrong.  No such
attitude\xspace map can exist on the bundle with $E = S^{2} \times S^{1}$,
so this bundle is not an adequate model for the space on which
spin-weighted functions should be defined.

\subsection{The Hopf bundle and the spin group}
\label{sec:hopf-bundle-spin}

We have shown that the naive choice of the trivial fiber bundle is not
sufficient.  Fortunately, there is another well known example of an
$S^{1}$ bundle over $S^{2}$: the Hopf bundle.\cite{Hopf1931,
  Hatcher2001, Vassiliev2001, Lyons2003, Urbantke2003} The total space
of this bundle is $E=S^{3}$.  Not only is this the appropriate
representation, allowing us to directly define simple functions for
both the bundle projection map and the attitude\xspace map, it will also
give us a clear geometric picture of the relationships between the
spaces.%
\footnote{It is useful to show explicitly that $S^{2} \times S^{1}$
  and $S^{3}$ are indeed topologically distinct spaces.  To see why,
  we can look at the fundamental groups of the spheres.  We have
  $\pi_{1}(S^{1}) \cong \mathbb{Z}$, and for $n \neq 1$ we have
  $\pi_{1}(S^{n}) \cong \unexpanded{\langle 0 \rangle}$, which is
  the trivial group consisting of just the identity.  Proposition 4.2
  of Ref.~\onlinecite{Hatcher2001} allows us to calculate
  $\pi_{1}(S^{2} \times S^{1}) \cong \pi_{1}(S^{2}) \times
  \pi_{1}(S^{1}) \cong \unexpanded{\langle 0 \rangle} \times
  \mathbb{Z} \cong \mathbb{Z}$, whereas
  $\pi_{1}(S^{3}) \cong \unexpanded{\langle 0 \rangle}$.  Since the
  fundamental group is a topological invariant, this shows that the
  two spaces cannot be homeomorphic.  As a historical note, it is
  significant that this conclusion also follows from considering
  $\pi_{2}$, but not any higher homotopy groups.  In particular
  $\pi_{3}(S^{2}) \cong \mathbb{Z}$---the proof of which was the
  surprising result from Hopf's original introduction of this fiber
  bundle.\cite{Hopf1931, Hatcher2001}} %

We begin with the traditional presentation of the Hopf map, by
expressing the various spheres in terms of their standard embeddings
into Cartesian space of one additional dimension:
$S^{1} \subset \mathbb{R}^{2}$, and so on.  Now, if a point on
$S^{3} \subset \mathbb{R}^{4}$ has Cartesian coordinates
$(w, x, y, z)$, the Hopf map $\mathfrak{h}: S^{3} \to S^{2}$ is defined
by\footnote{In Hopf's original presentation,\cite{Hopf1931} the order
  of the coordinates was slightly different; his form can be obtained
  from this one by swapping the $x$ and $z$ coordinates of $S^{3}$.
  The present ordering is adopted so that the quaternion mapping shown
  later will be more suited to other standard conventions.}
\begin{equation}
  \label{eq:HopfMap}
  \mathfrak{h}(w, x, y, z) = (
  2\,w\,y + 2\,x\,z,
  2\,y\,z - 2\,w\,x,
  w^{2} - x^{2} - y^{2} + z^{2}),
\end{equation}
where the right-hand side gives the point in Cartesian coordinates of
$\mathbb{R}^{3}$.  This is the projection map of the fiber bundle.
Some straightforward but wholly unenlightening algebra can be
used\cite{Hopf1931} to show directly that the preimage is indeed
homeomorphic to $S^{1}$; we will see this in a more elegant and
enlightening way below.

More interestingly, we can treat points in
$S^{3} \subset \mathbb{R}^{4}$ as unit quaternions, and points in
$S^{2} \subset \mathbb{R}^{3}$ as unit vectors.  The unit quaternions
form a group: the spin group $\mathrm{Spin}(3) \cong \mathrm{SU}(2)$.  A unit
quaternion $\mathbf{\MakeUppercase{R}}$ acts by conjugation on the unit vector
$\bm{z}$ in $\mathbb{R}^{3}$, using quaternion multiplication:
\begin{equation}
  \label{eq:rotor-action}
  \bm{v} = \mathbf{\MakeUppercase{R}}\, \bm{z}\, \mathbf{\MakeUppercase{R}}^{-1}.
\end{equation}
Here and below, we implicitly map between vectors and pure-vector
quaternions where needed by adding or removing a scalar component $0$.
If $\mathbf{\MakeUppercase{R}}$ has components $(w, x, y, z)$, then $\bm{v}$
has components given by Eq.~\eqref{eq:HopfMap}.  That is, rotation of
$\bm{z}$ by $\mathbf{\MakeUppercase{R}}$ is another expression of the Hopf
map:
\begin{equation}
  \label{eq:HopfMap-quaternions}
  \mathfrak{h}(\mathbf{\MakeUppercase{R}}) = \mathbf{\MakeUppercase{R}}\, \bm{z}\,
  \mathbf{\MakeUppercase{R}}^{-1}.
\end{equation}
This quaternionic presentation of the Hopf map has other nice
features: it allows us to explicitly calculate the fiber of any point,
and it is very closely related to the attitude\xspace map.

To calculate the fibers, we first find a single element of the fiber
over each point.  For each $\bm{v} \neq -\bm{z}$, we
map
\begin{subequations}
  \label{eq:fiber-elements}
  \begin{gather}
    \label{eq:fiber-elements-1}
    \bm{v} \mapsto \frac{1 - \bm{v} \, \bm{z}} {\sqrt{2 +
        2\, \bm{v} \cdot \bm{z}}},
    \intertext{and for the remaining point we arbitrarily choose}
    \label{eq:fiber-elements-2}
    -\bm{z} \mapsto \e^{\pi\, \bm{x}/2}.
  \end{gather}
\end{subequations}
It is not hard to show that the results are unit quaternions, and
correctly transform $\bm{z}$ into the desired vector in each case.
That is to say that they are indeed elements of the fiber over the
respective points.  Now given a single point in the fiber, we can find
all other points as follows.

Assuming $\mathbf{\MakeUppercase{R}}$ and $\mathbf{\MakeUppercase{R}}'$ represent two unit
quaternions in the fiber over a point, we know by definition that they
map to the same point on the sphere,
\begin{equation}
  \label{eq:fiber-condition}
  \mathbf{\MakeUppercase{R}}\, \bm{z}\, \mathbf{\MakeUppercase{R}}^{-1}
  = \mathbf{\MakeUppercase{R}}'\, \bm{z}\, \mathbf{\MakeUppercase{R}}'^{-1}.
\end{equation}
Intuitively, we would expect that the rotations represented by these
quaternions can only differ by an initial rotation about $\bm{z}$.
To see this more rigorously, we define
$\mathbf{\MakeUppercase{S}} \coloneqq \mathbf{\MakeUppercase{R}}^{-1}\, \mathbf{\MakeUppercase{R}}'$ and
rearrange Eq.~\eqref{eq:fiber-condition} to show that
$\mathbf{\MakeUppercase{S}}\, \bm{z} = \bm{z}\, \mathbf{\MakeUppercase{S}}$.  That is,
$\mathbf{\MakeUppercase{S}}$ commutes with $\bm{z}$.  The only quaternions
with this property are linear combinations of scalars and elements
proportional to $\bm{z}$.  Furthermore, $\mathbf{\MakeUppercase{S}}$ has unit
norm, as we can see from its definition.  All unit quaternions can be
written in the form
$\e^{\theta\, \bm{u}/2} = \cos \frac{\theta}{2} + \bm{u} \sin
\frac{\theta}{2}$, for some unit vector $\bm{u}$ and some scalar
$\theta$.\cite{Doran2010} Applied to this situation, that means
$\mathbf{\MakeUppercase{S}}$ must be of the form
$\mathbf{\MakeUppercase{S}} = \e^{\gamma\, \bm{z}/2}$ for some real number
$\gamma$; equivalently, we must have
$\mathbf{\MakeUppercase{R}}' = \mathbf{\MakeUppercase{R}}\, \e^{\gamma\, \bm{z}/2}$.
Obviously the exponential is periodic in $\gamma$ with period $4\pi$
(though the \emph{action} of this quaternion rotating any vector has
period $2\pi$).  Meanwhile, $\mathbf{\MakeUppercase{R}}$ can be any unit
quaternion taking $\bm{z}$ onto the point of $S^{2}$ in question.
Thus, we can calculate the fiber of any point on the sphere.
\begin{subequations}
  For $\bm{v} \neq -\bm{z}$, we have
  \label{eq:Hopf-fibers}
  \begin{gather}
    \mathfrak{h}^{-1}(\{\bm{v}\}) = \Set{ \frac{1 - \bm{v} \,
        \bm{z}} {\sqrt{2 + 2\, \bm{v} \cdot \bm{z}}}\,
      \e^{\gamma\, \bm{z}/2} | \gamma \in \mathbb{R}},
    \intertext{and we have} \mathfrak{h}^{-1}(\{-\bm{z}\}) = \Set{
      \e^{\pi\, \bm{x}/2}\, \e^{\gamma\, \bm{z}/2} | \gamma
      \in \mathbb{R}}.
  \end{gather}
\end{subequations}
By the periodicity in $\gamma$, we see that each such fiber is indeed
homeomorphic to $S^{1}$.

Now, because each fiber is topologically $S^{1}$, it can be used to
represent a unique choice of direction in the tangent space.  However,
it remains to be seen whether or not this can be done in a
\emph{continuous} way on the entire space $E = S^{3}$.  That is, we
need to show that there exists an attitude\xspace map $\mathfrak{a}$ as in
Eqs.~\eqref{eq:attitude-map}.  We define
\begin{equation}
  \label{eq:attitude-map-quaternions}
  \mathfrak{a}(\mathbf{\MakeUppercase{R}}) = \mathbf{\MakeUppercase{R}}\, \bm{x}\,
  \mathbf{\MakeUppercase{R}}^{-1}.
\end{equation}
The resulting vector is orthogonal to
$\bm{v} = \mathbf{\MakeUppercase{R}}\, \bm{z}\, \mathbf{\MakeUppercase{R}}^{-1}$, and
thus can be considered to be an element of the tangent space to the
sphere at that point.  The right-hand side of
Eq.~\eqref{eq:attitude-map-quaternions} is a rational polynomial in
the components of $\mathbf{\MakeUppercase{R}}$, and so is continuous everywhere
the denominator does not go to $0$, which is everywhere
$\mathbf{\MakeUppercase{R}} \neq 0$.  Of course, we assumed that $\mathbf{\MakeUppercase{R}}$
must have unit norm, so this condition is always satisfied, which
means that we have found an acceptable attitude\xspace map.\footnote{This
  construction is essentially the same as the ``flagpole'' description
  of spinors,\cite{Penrose1987} where $\mathfrak{h}$ provides the pole and
  $\mathfrak{a}$ provides the flag.}

We can evaluate this map on any element of the same fiber to find
\begin{subequations}
  \label{eq:attitude-rotated}
  \begin{align}
    \mathfrak{a}(\mathbf{\MakeUppercase{R}}\, \e^{\gamma\, \bm{z}/2})
    &= \mathbf{\MakeUppercase{R}}\, \e^{\gamma\, \bm{z}/2}\, \bm{x}\,
      \e^{-\gamma\, \bm{z}/2}\, \mathbf{\MakeUppercase{R}}^{-1}, \\
    &= \e^{\gamma\, \bm{v}/2}\, \mathbf{\MakeUppercase{R}}\, \bm{x}\,
      \mathbf{\MakeUppercase{R}}^{-1}\, \e^{-\gamma\, \bm{v}/2}, \\
    &= \e^{\gamma\, \bm{v}/2}\, \mathfrak{a}(\mathbf{\MakeUppercase{R}})\,
      \e^{-\gamma\, \bm{v}/2}.
  \end{align}
\end{subequations}
The final form of this expression shows that it is just the rotation
through an angle $\gamma$ about $\bm{v}$---in other words, it is a
rotation of the tangent space as we would expect.  Having exhibited
the existence of the map $\mathfrak{a}$, the Hairy-Ball Theorem now
tells us that there does not exist any \emph{global} section
$\mathfrak{s}$ of the Hopf bundle.  The map given by
Eqs.~\eqref{eq:fiber-elements} only constitutes a \emph{local} section
because it is discontinuous at $-\bm{z}$.  Thus, we avoid the
contradiction found above for $E = S^{2} \times S^{1}$.

The unit quaternions provided us with a simple realization of both the
Hopf bundle and the attitude\xspace map, as well as a clear geometric
picture of the relationship between the various spaces.  This suggests
that the group of unit quaternions---the spin group $\mathrm{Spin}(3)$---is,
in some fundamental way, the appropriate domain on which to define
spin-weighted spherical functions.  There is, however, a minor
subtlety in the attitude\xspace map, which might suggest a slightly
different domain, as we see next.

\subsection{Degree of the \texorpdfstring{attitude\xspace}{attitude} map and
  the domain of spin-weighted spherical functions}
\label{sec:degree-hopf-map}

In the above, an important feature of the projection and attitude\xspace
maps can be seen, but was not discussed explicitly.  Both $\mathfrak{h}$
and $\mathfrak{a}$ [Eqs.~\eqref{eq:HopfMap-quaternions}
and~\eqref{eq:attitude-map-quaternions}] involve $\mathbf{\MakeUppercase{R}}$
quadratically, meaning that $\mathbf{\MakeUppercase{R}}$ and $-\mathbf{\MakeUppercase{R}}$ map
to the same elements under both maps.  Since we have constructed these
maps as rotations of vectors, the interpretation is clear: the
$\mathrm{Spin}(3)$ group we have used is implicitly projected down to
$\mathrm{SO}(3)$, and the former is a double cover of the latter.  We can also
view this from another perspective.  Restricting attention to a single
fiber, the attitude\xspace map $\mathfrak{a}$ takes the fiber $S^{1}$ to the
space of unit tangent vectors, which is also homeomorphic to $S^{1}$.
But this map has degree $2$.  That is, the tangent vector
$\mathfrak{a}(\mathbf{\MakeUppercase{R}}\, \e^{\gamma\, \bm{z}/2})$ rotates
twice as $\gamma$ goes from $0$ to $4\pi$, even though these different
values of $\gamma$ all correspond to distinct quaternions on the fiber
before returning back to the starting point at $\gamma=4\pi$.  The
reason for this strange behavior is that we are considering tangent
vectors manipulated by spinors; vectors have spin $1$, whereas spinors
have spin $1/2$.

We can, if we wish, remove this strange feature by defining another
fiber bundle with this redundancy removed.  Here, the entire space is
just the projective sphere $\mathbb{R}\mathrm{P}^{3}$, which is equivalent to the sphere
$S^{3}$ with antipodal points identified.  This is naturally the
topology of $\mathrm{SO}(3)$, so we can identify each point in $\mathbb{R}\mathrm{P}^{3}$ with
an operator in $\mathrm{SO}(3)$.  Then, we can again form a projection map
$\mathfrak{p} : \mathbb{R}\mathrm{P}^{3} \to S^{2}$ by taking
$\mathfrak{p}(\mathcal{R}) = \mathcal{R}(\bm{z})$, which is just
the vector $\bm{z}$ rotated by $\mathcal{R} \in \mathrm{SO}(3)$.
Similarly, we can define the attitude\xspace map as
$\mathfrak{a}(\mathcal{R}) = \mathcal{R}(\bm{x})$.  This is nearly
the same construction as above,\footnote{Again, it is useful to see
  that this truly is a distinct fiber bundle over the sphere.  In this
  case, we merely need to show that $S^{3}$ and $\mathbb{R}\mathrm{P}^{3}$ are
  topologically distinct.  But this is already known from the familiar
  fact that $\mathrm{Spin}(3) \cong \mathrm{SU}(2)$ is simply connected, whereas
  $\mathrm{SO}(3)$ is not.\cite{Hatcher2001, Frankel2004}  In fact, with this
  construction, we have demonstrated that the Hopf bundle provides a
  ``spin structure'' over the bundle $\mathrm{SO}(3)$.\cite{Milnor1963}}
except that this attitude\xspace map has degree $1$.

However, this projective construction is somewhat complicated.  And
there is no particular reason to avoid the original formulation in
terms of $\mathrm{Spin}(3)$; we can still use it to construct functions of
integer spin weight in a natural way, even with this small amount of
redundancy.  More importantly, we can also use $\mathrm{Spin}(3)$ to construct
functions of half-integer spin weight---for which we cannot use
$\mathrm{SO}(3)$.  Finally, as a simple practical matter, parametrizations of
$\mathrm{SO}(3)$ either are just parametrizations of $\mathrm{Spin}(3)$ with this
redundancy present, or can be extended trivially to parametrizations
of $\mathrm{Spin}(3)$---as will be discussed further in
Appendix~\ref{sec:param-spher}.  Taken together, these arguments
indicate that there is no good reason to restrict the domain to
$\mathrm{SO}(3)$, and every reason to use $\mathrm{Spin}(3)$ as the domain instead.

\section{The codomain of SWSFs}
\label{sec:codomain}

Now, having settled on the appropriate domain for SWSFs, we need to
understand the codomain in which these functions will take values.
Traditional presentations\cite{Newman1966, Goldberg1967,
  Eastwood1982, Penrose1987} chose the complex numbers $\mathbb{C}$ as
the codomain.  It should be noted that the spinor space over any
two-dimensional vector space with Euclidean norm is---algebraically
speaking---identical to the complex
numbers,\cite{francis_construction_2005} but the geometric
interpretation of this codomain is not clear.  An intuitively obvious
choice would be the vector space of the $\bm{a}$-$\bm{b}$
plane, where
\begin{subequations}
  \label{eq:tangent-plane}
  \begin{align}
    \bm{a} &\coloneqq \mathbf{\MakeUppercase{R}}\, \bm{x}\,
    \mathbf{\MakeUppercase{R}}^{-1}, \\
    \bm{b} &\coloneqq \mathbf{\MakeUppercase{R}}\, \bm{y}\,
    \mathbf{\MakeUppercase{R}}^{-1}.
  \end{align}
\end{subequations}
If we think of the set of all directions
$\bm{n} \coloneqq \mathbf{\MakeUppercase{R}}\, \bm{z}\, \mathbf{\MakeUppercase{R}}^{-1}$ as
comprising a sphere $S^{2}$ embedded in $\mathbb{R}^{3}$, the
$\bm{a}$-$\bm{b}$ plane is just the tangent plane at $\bm{n}$.
This would be in line with our original motivation for SWSFs as
representing the component of some radiated or received wave.  This
interpretation is technically slightly complicated because the
codomain in this case would be the spinor space of the entire
three-dimensional vector space, but the function on $\mathbf{\MakeUppercase{R}}$
would only take values in the spinor subalgebra corresponding to the
plane spanned by the vectors of Eqs.~\eqref{eq:tangent-plane}.

Alternatively, $\mathbb{C}$ could correspond to the spinor space of
the $\bm{x}$-$\bm{y}$ plane, independent of the argument of
the function.  This is a somewhat tidier choice, as the codomain is
``minimal'' in some sense.  Geometrically, the
$\bm{x}$-$\bm{y}$ plane can be rotated onto the
$\bm{a}$-$\bm{b}$ plane by $\mathbf{\MakeUppercase{R}}$.  Algebraically,
the spinor is rotated by conjugation by $\mathbf{\MakeUppercase{R}}$, just as the
vectors are in Eqs.~\eqref{eq:tangent-plane}.  So we could use the
same \emph{interpretation} of the spinors as describing the
$\bm{a}$-$\bm{b}$ plane, even though the actual function
values are in the arbitrarily chosen $\bm{x}$-$\bm{y}$ plane.
We will see in Sec.~\ref{sec:natural-basis} that this is actually the
standard choice, and is covariant in the sense that we can transform
any choice of $\bm{x}$-$\bm{y}$ plane into any other and hence
also transform the codomain in a consistent fashion.  As usual, when
using a more geometric interpretation like this, the unit imaginary
$\i$ that is found in standard treatments of spherical harmonics and
angular momentum, and was used freely in
Sec.~\ref{sec:spin-weight-funct}, will be replaced by a geometric
object---in this case, the bivector representing the
$\bm{x}$-$\bm{y}$ plane.

The codomain Straumann\cite{Straumann2014} chose appears to be unique
in the literature: for integer-weight SWSHs, he chose the codomain to
be a two-dimensional subspace $\mathcal{M}$ of the Lie algebra
$\mathfrak{so}(3) = \mathcal{M} \oplus \mathfrak{so}(2)$, where $\mathfrak{so}(2)$ is the algebra
corresponding to the fiber space $S^{1}$.  In general the geometry
behind this choice is not specified.  However, when Straumann
specializes to Euler angle coordinates, the $\mathfrak{so}(2)$ fiber corresponds
to an initial rotation about $\bm{z}$, so that $\mathcal{M}$
constitutes generators of rotations about $\bm{x}$ and
$\bm{y}$.  So in this presentation SWSHs take values in the space
of vectors in the $\bm{x}$-$\bm{y}$ plane.  This use of
vectors instead of spinors is important to Straumann's approach,
because it allows him to use familiar constructions in differential
geometry.  Of course, it is well known that spinor and vector
representations of the plane are equivalent for quantities of integer
spin;\cite{Doran2010} the underlying geometry of the codomain is the
same in either case.

\section{A natural basis for SWSFs}
\label{sec:natural-basis}

We have determined that SWSFs are appropriately defined as functions
from $\mathrm{Spin}(3)$ to the spinor algebra of two dimensions---familiar as
the complex numbers $\mathbb{C}$.  The latter of these has various
reasonable geometric interpretations, but the standard one involves
the $\bm{x}$-$\bm{y}$ plane.  In fact,
Wigner\cite{Wigner1959} introduced a canonical set of functions from
$\mathrm{Spin}(3)$ to $\mathbb{C}$, which are known as Wigner's $\ensuremath{\mathfrak{D}} $ matrices.
Using the elements of these matrices, the standard SWSHs may now be
redefined as
\begin{equation}
  \label{eq:SWSH-definition}
  \ensuremath{\scripts{_{s}}{Y}{_{\ell,m}}}(\mathbf{\MakeUppercase{R}}) \coloneqq (-1)^{s} \sqrt{ \frac{2\ell +
      1} {4\pi}} \ensuremath{\mathfrak{D}} ^{(\ell)}_{m, -s} (\mathbf{\MakeUppercase{R}}),
\end{equation}
which is a slight extension of a formula already known to Goldberg
\etal,\cite{Goldberg1967} but here defining the SWSHs on $\mathrm{Spin}(3)$,
rather than on coordinate systems of $S^{2}$.  (Also note that the
factor $(-1)^{s}$ differs from that of Goldberg \etal, but is
consistent with more modern standards.\cite{Ajith2007b})  Moreover,
Bargmann\cite{Bargmann1947} and Gelfand, Minlos, and
Shapiro\cite{Gelfand1963} proved that this collection of
functions---encompassing all possible $\ell$, $m$, and $s$
values---forms a complete orthogonal system in the space of
square-integrable complex-valued functions on $S^{3}$.  In this
section we will first use Eq.~\eqref{eq:SWSH-definition} to briefly
examine one benefit of defining SWSHs in this way, on the full spin
group.  We will see that these SWSHs transform among themselves, which
cannot be said for the more standard SWSHs defined on coordinates of
$S^{2}$.  We will then derive the $\ensuremath{\mathfrak{D}} $ matrices using the geometric
structure established by $\mathrm{Spin}(3)$, being careful about special
cases, and suggesting improved methods for evaluating the necessary
sums with greater numerical efficiency and accuracy.  Finally, we will
review the geometric features of SWSHs defined in this way, connecting
back to the motivation mentioned in Sec.~\ref{sec:Introduction}
relating to waves measured on a sphere.

\subsection{Transforming SWSHs}
\label{sec:transforming-swshs}
Wigner constructed his $\ensuremath{\mathfrak{D}} ^{(\ell)}$ matrices to form a representation
of the spin group.  That is, given $\mathbf{\MakeUppercase{R}}_{1}$ and
$\mathbf{\MakeUppercase{R}}_{2}$ in $\mathrm{Spin}(3)$, we have
\begin{equation}
  \label{eq:WignerD-representation}
  \ensuremath{\mathfrak{D}} ^{(\ell)}_{m',m}(\mathbf{\MakeUppercase{R}}_{1}\, \mathbf{\MakeUppercase{R}}_{2})
  = 
  \sum_{m''} \ensuremath{\mathfrak{D}} ^{(\ell)}_{m',m''}(\mathbf{\MakeUppercase{R}}_{1})\, 
  \ensuremath{\mathfrak{D}} ^{(\ell)}_{m'',m}(\mathbf{\MakeUppercase{R}}_{2}).
\end{equation}
This simple formula now allows us to relate SWSHs defined with respect
to different frames.  For example, if one frame is taken into the
other by some rotation $\mathbf{\MakeUppercase{R}}_{\text{f}}$ so that we can write
$\mathbf{\MakeUppercase{R}}_{1} = \mathbf{\MakeUppercase{R}}_{\text{f}}\, \mathbf{\MakeUppercase{R}}_{2}$, we have
\begin{equation}
  \label{eq:SWSH-transformation}
  \ensuremath{\scripts{_{s}}{Y}{_{\ell,m}}}(\mathbf{\MakeUppercase{R}}_{1}) = \sum_{m'} \ensuremath{\mathfrak{D}} ^{(\ell)}_{m, m'}(\mathbf{\MakeUppercase{R}}_{\text{f}})\,
  \ensuremath{\scripts{_{s}}{Y}{_{\ell,m'}}}(\mathbf{\MakeUppercase{R}}_{2}).
\end{equation}
The important feature of this equation is that the
$\ensuremath{\mathfrak{D}} ^{(\ell)}_{m, m'}$ factors are all constant.  That is, given $\mathbf{\MakeUppercase{R}}_{\text{f}}$,
these are complex coefficients independent of $\mathbf{\MakeUppercase{R}}_{1}$ and
$\mathbf{\MakeUppercase{R}}_{2}$.  So we can truly express
$\ensuremath{\scripts{_{s}}{Y}{_{\ell, m}}}(\mathbf{\MakeUppercase{R}}_{1})$ as a finite and closed-form
expansion in $\ensuremath{\scripts{_{s}}{Y}{_{\ell,m'}}}(\mathbf{\MakeUppercase{R}}_{2})$ for various values
of $m'$.  This is useful because it allows us to transform the value
of a function given with respect to one frame into a different frame,
which is a necessary step in some types of transformations, as
explained in detail in Ref.~\onlinecite{Boyle2015}.

It is important to note that there is no comparable transformation law
when the $\ensuremath{\scripts{_{s}}{Y}{_{\ell,m}}}$ functions are defined as a function of
coordinates on the sphere $S^{2}$.  Specifically, we have
\begin{equation}
  \label{eq:SWSH-bad-transformation}
  \ensuremath{\scripts{_{s}}{Y}{_{\ell,m}}}(\vartheta_{1}, \varphi_{1}) = \sum_{m'}
  \ensuremath{\mathfrak{D}} ^{(\ell)}_{m, m'}(\mathbf{\MakeUppercase{R}}_{\text{f}})\, \e^{-\i\, s\, \gamma}\,
  \ensuremath{\scripts{_{s}}{Y}{_{\ell,m'}}}(\vartheta_{2}, \varphi_{2}).
\end{equation}
But here, the angle $\gamma$ is a function of both $\mathbf{\MakeUppercase{R}}_{\text{f}}$ and
$(\vartheta_{2}, \varphi_{2})$.  Moreover, for any nontrivial rotation
$\mathbf{\MakeUppercase{R}}_{\text{f}}$ this function $\gamma$ is very complicated---even when $\mathbf{\MakeUppercase{R}}_{\text{f}}$ is
a simple rotation about any special axis, such as $\bm{x}$ or $\bm{y}$.
In fact, we can calculate this angle using the definitions of the
coordinate systems.  We have
$\e^{\varphi_{1}\, \bm{z}/2}\, \e^{\vartheta_{1}\, \bm{y}/2} =
\mathbf{\MakeUppercase{R}}_{\text{f}}\, \e^{\varphi_{2}\, \bm{z}/2}\, \e^{\vartheta_{2}\,
  \bm{y}/2}\, \e^{\gamma\, \bm{z}/2}$.  The final factor is
necessary because the $(\vartheta_{i}, \varphi_{i})$ rotations are
restricted to special forms.  Solving for $\gamma$, we get
\begin{equation}
  \label{eq:SWSH-gamma-calculation}
  \gamma = -2\, \bm{z}\, \log \left[\e^{-\vartheta_{2}\, \bm{y}/2}\,
    \e^{-\varphi_{2}\, \bm{z}/2}\, \mathbf{\MakeUppercase{R}}_{\text{f}}^{-1}\, \e^{\varphi_{1}\,
      \bm{z}/2}\, \e^{\vartheta_{1}\, \bm{y}/2}\, \right].
\end{equation}
Of course, $(\vartheta_{1}, \varphi_{1})$ and
$(\vartheta_{2}, \varphi_{2})$ are already related in a very
complicated way when $\mathbf{\MakeUppercase{R}}_{\text{f}}$ is anything other than a pure rotation
about $\bm{z}$.  An equivalent result holds, of course, when SWSHs
are defined on stereographic coordinates.  In any case, $\gamma$ is a
very complicated function of the coordinates---not just of the
rotation.  Because of this $\gamma$ function, it cannot be said that
the $\ensuremath{\scripts{_{s}}{Y}{_{\ell,m}}}$ functions, when defined on coordinates of $S^{2}$,
transform among themselves.

To be fair, we should note that while the SWSHs do not transform among
themselves in this form, the \emph{modes} of a spin-weighted function
decomposed into SWSHs do transform among themselves under rotation.
That is, if we have
\begin{equation}
  \label{eq:decomposition}
  \prefixscripts{_{s}}{f}(\vartheta_{1}, \varphi_{1}) = \sum_{\ell,m}
  f_{1}^{\ell,m} \ensuremath{\scripts{_{s}}{Y}{_{\ell,m}}}(\vartheta_{1},
  \varphi_{1})
\end{equation}
for some constants $f_{1}^{\ell,m}$, then we can use
Eqs.~\eqref{eq:SWSH-bad-transformation} and~\eqref{eq:spin-weight} to
see that the equivalent constants $f_{2}^{\ell,m}$ defined with
respect to a different basis are related by
\begin{equation}
  \label{eq:mode-transformation}
  f_{2}^{\ell,m} = \sum_{m'} f_{1}^{\ell,m'} \ensuremath{\mathfrak{D}} ^{(\ell)}_{m', m}(\mathbf{\MakeUppercase{R}}_{\text{f}}).
\end{equation}
Note that this relationship depends only on the rotation $\mathbf{\MakeUppercase{R}}_{\text{f}}$, and
not on any coordinates.  Naturally, the fact that the modes transform
among themselves is extremely useful when treating only rotations.
Unfortunately, it is not true when the transformation is more
complicated---as with general Lorentz transformations.\cite{Boyle2015}

\subsection{Defining Wigner's \texorpdfstring{$\ensuremath{\mathfrak{D}} $}{D} functions}
\label{sec:defining-wigners-d}
As explained in Sec.~\ref{sec:codomain}, our SWSH functions will map
into the spinor subalgebra of the $\bm{x}$-$\bm{y}$ plane.
This subalgebra consists of linear combinations of the quaternion $1$
and the quaternion representing the $\bm{x}$-$\bm{y}$
plane---which is actually $\bm{z}$.\footnote{It turns out that the
  ``vector'' part of a quaternion would more accurately be described
  as a ``bivector'' part; we have simply followed tradition and used
  the same notation for vectors and bivectors.  This is possible only
  by sheer coincidence and the fact that we work in three dimensions.
  This coincidence and the consequent confusion led to the
  quaternion-vector wars of the late nineteenth
  century,\cite{Crowe1985} which may be the greatest mathematical
  tragedy in the history of physics.  For our purposes, the important
  point is that the $\bm{x}$-$\bm{y}$ plane is represented by
  the bivector $\bm{x} \wedge \bm{y}$, which is traditionally
  denoted $\bm{z}$.}  Our first task will be to decompose the SWSH
function's argument $\mathbf{\MakeUppercase{R}} \in \mathrm{Spin}(3)$ into two parts, each
of which will be an element of this spinor subalgebra.  We do this by
taking parts of $\mathbf{\MakeUppercase{R}}$ that are symmetric (do not change)
and antisymmetric (change sign) under rotation by $\pi$ about the $z$
axis.

We can express the geometric notion of this rotation by the algebraic
notion of conjugation by $\bm{z}$---that is, multiplying on the left by
$\bm{z}$ and on the right by $\bm{z}^{-1}$.\footnote{A full understanding of
  the geometry behind these operations requires Geometric Algebra
  (GA)\cite{Hestenes1987, Hestenes2002, Doran2010} which is,
  mathematically speaking, just Clifford Algebra (CA) over the field
  $\mathbb{R}$.  GA allows us to replace essentially all uses of
  complex numbers, quaternions, and more in physics (including quantum
  mechanics) with geometric constructions involving only real
  numbers.\cite{Doran2010} Moreover, through numerous publications
  over decades of work, Hestenes\cite{Hestenes1987, Hestenes2002} has
  provided GA with additional geometric interpretation that moves
  beyond the mere formalism of CA.  Though GA is a surprisingly
  elementary subject, explaining its relevance here would take us
  slightly beyond the scope of this paper, and yield little more of
  direct relevance than the basic complex and quaternion
  formalisms---which are just the first two nontrivial examples of
  GA.}  The symmetric part of $\mathbf{\MakeUppercase{R}}$ will be a linear
combination of $1$ and $\bm{z}$, while the antisymmetric part will be a
linear combination of $\bm{x}$ and $\bm{y}$.  The latter can then be mapped
into the same space as the former by multiplying on the left by
$\bm{y}^{-1}$, giving us two objects in the same two-dimensional spinor
subalgebra---complex numbers, but with a geometric interpretation.
Explicitly, we define
\begin{subequations}
  \label{eq:quaternion-parts}
  \begin{align}
    R_{\text{s}} &\coloneqq \frac{1}{2} \left( \mathbf{\MakeUppercase{R}} + \bm{z}\,
          \mathbf{\MakeUppercase{R}}\, \bm{z}^{-1} \right), \\
    R_{\text{a}} &\coloneqq \frac{1}{2}\, \bm{y}^{-1} \left( \mathbf{\MakeUppercase{R}} - \bm{z}\,
          \mathbf{\MakeUppercase{R}}\, \bm{z}^{-1} \right).
  \end{align}
\end{subequations}
In terms of components, if
$\mathbf{\MakeUppercase{R}} = R_{1} 1 + R_{\bm{x}}\, \bm{x} + R_{\bm{y}}\, \bm{y} + R_{\bm{z}}\,
\bm{z}$, we have
\begin{subequations}
  \label{eq:quaternion-parts-components}
  \begin{align}
    R_{\text{s}} &= R_{1} + R_{\bm{z}}\, \bm{z}, \\
    R_{\text{a}} &= R_{\bm{y}} + R_{\bm{x}}\, \bm{z} .
  \end{align}
\end{subequations}
Noting that the coefficients are real numbers, while $\bm{z}^{2} = -1$,
these quantities act precisely like complex numbers.  This
decomposition obeys an important product law: for any other quaternion
$\mathbf{\MakeUppercase{S}}$, we have
\begin{subequations}
  \label{eq:product-law}
  \begin{align}
    \left( \mathbf{\MakeUppercase{R}}\, \mathbf{\MakeUppercase{S}} \right)_{\text{s}}
    &= R_{\text{s}}\, S_{\text{s}} - \bar{R}_{\text{a}}\, S_{\text{a}}, \\
    \left( \mathbf{\MakeUppercase{R}}\, \mathbf{\MakeUppercase{S}} \right)_{\text{a}}
    &= R_{\text{a}}\, S_{\text{s}} + \bar{R}_{\text{s}}\, S_{\text{a}}.
  \end{align}
\end{subequations}
We could, of course, accomplish a similar decomposition using any two
orthogonal unit ``pure-imaginary'' quaternions in place of $\bm{z}$ and
$\bm{y}$.  The particular choices of Eq.~\eqref{eq:quaternion-parts} are
made to correspond more directly with conventional presentations
elsewhere.  However, this choice must satisfy one important
constraint: that the right-hand sides of Eq.~\eqref{eq:product-law}
are linear combinations of $S_{\text{a}}$ and $S_{\text{s}}$, rather than their complex
conjugates.  Furthermore, these equations should reduce to identities
when either $\mathbf{\MakeUppercase{R}}$ or $\mathbf{\MakeUppercase{S}}$ is $1$.

The derivation proceeds from here by constructing a
$(2\ell+1)$-dimensional vector space consisting of these spinors,
where $\ell$ can be any non-negative integer or half-integer.
Following Wigner,\cite{Wigner1959} we can do this by providing a
basis explicitly:
\begin{equation}
  \label{eq:D-basis}
  \mathbf{e}_{(m)} ( \mathbf{\MakeUppercase{S}} ) \coloneqq \frac{S_{\text{s}}^{\ell+m}\,
    S_{\text{a}}^{\ell-m}} {\sqrt{ (\ell+m)!\, (\ell-m)! }}.
\end{equation}
As usual, $m$ varies from $-\ell$ to $\ell$ in integer steps.  We can
also replace $\mathbf{\MakeUppercase{S}}$ with $\mathbf{\MakeUppercase{R}}\, \mathbf{\MakeUppercase{S}}$
in this expression.  The result can be expanded in terms of the
original basis given here.  We then define the $\ensuremath{\mathfrak{D}} $ matrix as the
relevant expansion coefficients:
\begin{equation}
  \label{eq:WignerD-definition}
  \mathbf{e}_{(m')} ( \mathbf{\MakeUppercase{R}}\, \mathbf{\MakeUppercase{S}} ) = \sum_{m'}
  \ensuremath{\mathfrak{D}} ^{(\ell)}_{m',m} (\mathbf{\MakeUppercase{R}})\, \mathbf{e}_{(m)} (
  \mathbf{\MakeUppercase{S}} ).
\end{equation}
We can expand the left-hand side here by inserting the right-hand
sides of Eq.~\eqref{eq:product-law} into the right-hand side of
Eq.~\eqref{eq:D-basis}.  Since $\ell \pm m$ is always a non-negative
integer, we can use the binomial theorem to expand each of the
factors, then group the resulting terms to find the expansion
coefficients $\ensuremath{\mathfrak{D}} ^{(\ell)}_{m',m}$ for this quantity.  The naive
calculation provides this expression, which (after accounting for
minor differences in conventions) is the same as the formula given by
Wigner:\cite{Wigner1959}
\begin{equation}
  \label{eq:WignerD-naive}
  \mathfrak{D}^{(\ell)}_{m',m}(\mathbf{\MakeUppercase{R}}) = \sum_{\rho}
  \binom{\ell+m'} {\rho}\, \binom{\ell-m'} {\ell-\rho-m}\,
  (-1)^{\rho}\, R_{\text{s}}^{\ell+m'-\rho}\, \bar{R}_{\text{s}}^{\ell-\rho-m}\,
  R_{\text{a}}^{\rho-m'+m}\, \bar{R}_{\text{a}}^{\rho}\, \sqrt{ \frac{ (\ell+m)!\,
      (\ell-m)! } { (\ell+m')!\, (\ell-m')! } }.
\end{equation}
The summation variable $\rho$ simply ranges over all values for which
the binomial coefficients are nonzero.  This expression is very
inefficient to calculate directly, and is subject to enormous errors
or even arithmetic overflow---in which some of the factors are too
large for computers to represent natively.  We can refine the
expression to be faster, more accurate, and deal with special cases
efficiently.

We introduce four branches to the calculation of $\ensuremath{\mathfrak{D}} $, depending on
the value of $\mathbf{\MakeUppercase{R}}$.  First, we deal with the cases where
either $\left\lvert{R_{\text{s}}}\right\rvert < \epsilon$ or $\left\lvert{R_{\text{a}}}\right\rvert < \epsilon$ for some small
number $\epsilon$ comparable to machine precision.  In either such
case, we can ignore the smaller quantity and the product
law~\eqref{eq:product-law} becomes simple.  Then, depending on which
component is smaller,
$\mathbf{e}_{(m')} ( \mathbf{\MakeUppercase{R}}\, \mathbf{\MakeUppercase{S}} )$ is simply
proportional to either $\mathbf{e}_{(m')} ( \mathbf{\MakeUppercase{S}} )$ or
$\mathbf{e}_{(-m')} ( \mathbf{\MakeUppercase{S}} )$.  If neither $\left\lvert{R_{\text{s}}}\right\rvert$ nor
$\left\lvert{R_{\text{a}}}\right\rvert$ is small, we must use Eq.~\eqref{eq:WignerD-naive}, but we
can extract the constant terms, and express it as a polynomial in some
constant.  Here again, we distinguish between two cases, where we use
the smaller of $\left\lvert{R_{\text{s}}/R_{\text{a}}}\right\rvert^{2}$ or $\left\lvert{R_{\text{a}}/R_{\text{s}}}\right\rvert^{2}$ as the
expansion variable in which to express the polynomial.  The polynomial
should be evaluated using a generalization of Horner form [see
Appendix~\ref{sec:eval-polyn-with} for more details] for improved
speed and accuracy.  Finally, the complex powers of the terms we
factor out in these cases must be evaluated in a polar decomposition
to avoid arithmetic overflow and to increase the speed of evaluation.
For this purpose, we define the auxiliary variables
\begin{subequations}
  \begin{align}
    r_{\text{s}} &\coloneqq \left\lvert{R_{\text{s}}}\right\rvert, &
    \phi_{\text{s}} &\coloneqq \arg{R_{\text{s}}}, \\
    r_{\text{a}} &\coloneqq \left\lvert{R_{\text{a}}}\right\rvert, &
    \phi_{\text{a}} &\coloneqq \arg{R_{\text{a}}}.
  \end{align}
\end{subequations}
Functions are available in many standard software libraries to obtain
both the modulus and argument simultaneously for increased accuracy
and speed, and to translate back from this polar decomposition to the
usual rectangular form of complex numbers.

Putting this all together, the result is that in practice it is best
to calculate the $\ensuremath{\mathfrak{D}} $ matrices according to
\begin{widetext}
  \begingroup \let\displaystyle\textstyle 
  \begin{equation}
    \label{eq:WignerD-general}
    \ensuremath{\mathfrak{D}} ^{(\ell)}_{m',m}(\mathbf{\MakeUppercase{R}}) =
    \begin{cases}
      \e^{2 \i m \phi_{\text{s}}}\, \delta_{m',m} & r_{\text{a}} < \epsilon,
      \\[1.5ex]
      (-1)^{\ell+m} \e^{2 \i m \phi_{\text{a}}}\, \delta_{-m',m} & r_{\text{s}} < \epsilon,
      \\[2ex]
      \!\begin{aligned} 
        &\sqrt{ \frac{(\ell+m)! (\ell-m)!} {(\ell+m')! (\ell-m')!} }
        r_{\text{s}}^{2\ell - m + m' - 2\rho_{1}} r_{\text{a}}^{m - m' + 2\rho_{1}}
        \\
        &\quad \times \e^{\i [\phi_{\text{s}}(m+m') + \phi_{\text{a}}(m-m')]}
        \sum_{\rho=\rho_{1}}^{\rho_{2}} \binom{\ell+m'} {\rho}
        \binom{\ell - m'} {\ell-m-\rho} \left( - r_{\text{a}}^{2} / r_{\text{s}}^{2}
        \right)^{\rho}
      \end{aligned}
      & r_{\text{a}} \leq r_{\text{s}},
      \\[5ex]
      \!\begin{aligned} 
        &\sqrt{ \frac{(\ell+m)!  (\ell-m)!} {(\ell+m')!  (\ell-m')!} }
        r_{\text{a}}^{2\ell - m - m' - 2\rho_{3}} r_{\text{s}}^{m + m' +
          2\rho_{3}} (-1)^{\ell+m} \\
        &\quad \times \e^{\i [\phi_{\text{a}}(m-m') + \phi_{\text{s}}(m+m')]}
        \sum_{\rho=\rho_{3}}^{\rho_{4}} \binom{\ell+m'} {\ell-m-\rho}
        \binom{\ell - m'} {\rho} \left( - r_{\text{s}}^{2} / r_{\text{a}}^{2}
        \right)^{\rho}
      \end{aligned}
      & r_{\text{s}} < r_{\text{a}}.
    \end{cases}
  \end{equation}
  \endgroup
\end{widetext} %
Again, this expression is valid for integer or half-integer $\ell$,
and the sums are evaluated more quickly and accurately using the
algorithm presented in Appendix~\ref{sec:eval-polyn-with}.  [Note that
$\bm{z}$ has been replaced with $\i$ here, so as to not confuse the
reader who has consulted this paper only for this equation.  In any
case, because they are algebraically identical, this form is likely to
be more similar to the form in which the equation should be
implemented in computer code.]  The limits of the sums are
\begin{subequations}
  \begin{align}
    \label{eq:WignerD-sum-limits}
    \rho_{1} &= \max(0, m'-m), \\
    \rho_{2} &= \min(\ell+m', \ell-m), \\
    \rho_{3} &= \max(0, -m'-m), \\
    \rho_{4} &= \min(\ell-m', \ell-m).
  \end{align}
\end{subequations}
Unfortunately, because the sum alternates in sign, this is still
numerically unstable for certain values of $\mathbf{\MakeUppercase{R}}$ and
$(\ell, m', m)$---specifically, for $r_{\text{s}} \approx r_{\text{a}}$ and
$m' \approx m \approx 0$ when $\ell \gtrsim 12$, as discussed further
in Appendix~\ref{sec:eval-polyn-with}.  If accurate values are needed
for such values, these expressions can be used to initialize recursion
relations,\cite{Choi1999} which are naturally independent of the
parametrization of $\ensuremath{\mathfrak{D}} $ and can accurately determine the values of
these elements with $m' \approx m \approx 0$.

\subsection{Geometric interpretation of SWSHs and SWSFs}
\label{sec:geom-interpr-result}

Now, equipped with a more detailed understanding of SWSFs, and a
particular realization in the form of SWSHs, we can return to the
issue of representing waves on a sphere $S^{2}$.  We began with an
arbitrary reference frame $(\bm{x}, \bm{y}, \bm{z})$.  Using any element
$\mathbf{\MakeUppercase{R}}$ in the spin group $\mathrm{Spin}(3)$, we defined another
frame
\begin{subequations}
  \label{eq:tangent-frame}
  \begin{align}
    \bm{a} &\coloneqq \mathbf{\MakeUppercase{R}}\, \bm{x}\,
    \mathbf{\MakeUppercase{R}}^{-1}, \\
    \bm{b} &\coloneqq \mathbf{\MakeUppercase{R}}\, \bm{y}\,
    \mathbf{\MakeUppercase{R}}^{-1}, \\
    \bm{n} &\coloneqq \mathbf{\MakeUppercase{R}}\, \bm{z}\,
    \mathbf{\MakeUppercase{R}}^{-1}.
  \end{align}
\end{subequations}
The space of all possible directions $\bm{n}$ maps out the sphere,
and the tangent plane at any such point is spanned by $\bm{a}$ and
$\bm{b}$.  Now, the first point to note is that a transverse wave
must be given by some vector in the space spanned by $\bm{a}$ and
$\bm{b}$, or some tensor or spinor constructed with those vectors.
But we've expressed the SWSHs above as functions mapping into the
spinor space of the $\bm{x}$-$\bm{y}$ plane---that is, linear
combinations of $1$ and $\bm{z}$.  So we may wonder how we can uniquely
relate quantities in one space to quantities in another.  Fortunately,
$\mathbf{\MakeUppercase{R}}$ gives us a solution to this problem: we simply rotate
the $\bm{x}$-$\bm{y}$ spinor via conjugation by $\mathbf{\MakeUppercase{R}}$, just
like with a vector.

An example will be helpful.  Gravitational waves are typically modeled
by a perturbation $h^{\mu \nu}$ to the metric tensor, where the
background metric is assumed to be Minkowski.  Furthermore, the gauge
can be chosen so that $h^{\mu \nu}$ is traceless, and the perturbation
is transverse to the direction in which the wave is
traveling.\cite{Misner1973} This is then conventionally combined into
a single complex number as\footnote{Clearly, under rotation of the
  tangent frame, $h$ transforms with spin weight $s=-2$.  Conversely,
  given $h$, we can reconstruct the perturbation as
  $h^{\mu \nu} = h\, m^{\mu}\, m^{\nu}$, which has no spin weight
  because it is a tensor.  The complex ``scalar'' quantity $h$ is able
  to contain all of this information because the gauge is chosen so
  that $h^{\mu \nu}$ is transverse, traceless, and symmetric.}
\begin{subequations}
  \label{eq:metric-perturbation}
  \begin{align}
    \label{eq:metric-perturbation1}
    h
    &\coloneqq \frac{1}{2} h^{\mu \nu} \left[ \left( a_{\mu} a_{\nu}
      - b_{\mu} b_{\nu} \right) - \i\, \left(a_{\mu} b_{\nu}
      + b_{\mu} a_{\nu}\right) \right], \\
    \label{eq:metric-perturbation2}
    &\equiv h^{\mu \nu} \bar{m}_{\mu}\, \bar{m}_{\nu}.
  \end{align}
\end{subequations}
In the second line we have used the vector $\bm{m}$ defined in
Eq.~\eqref{eq:mVector}.  The appearance of $\i$ explicitly in
Eq.~\eqref{eq:metric-perturbation1} and implicitly in
Eq.~\eqref{eq:metric-perturbation2} is conventional but not necessary;
as always, the unit imaginary is generally best replaced by something
with geometric significance.\cite{Gull1993}  We used a geometric
construction in Sec.~\ref{sec:defining-wigners-d} to avoid the
arbitrary introduction of the complex quantity $\i$, in favor of
$\bm{z}$.  In particular, the value of a SWSH will be an element of the
spinor space spanned by $1$ and $\bm{z}$.  But $\bm{z}$ has no direct
geometric significance for an arbitrary point on the sphere or its
tangent space.  The more natural construction here is to replace $\i$
with $\bm{n}$.  Now, if the value of the SWSH is $\alpha + \beta \bm{z}$
for some real numbers $\alpha$ and $\beta$, then we can write
\begin{equation}
  \label{eq:rotating-SWSH-spinor}
  \mathbf{\MakeUppercase{R}} (\alpha + \beta \bm{z}) \mathbf{\MakeUppercase{R}}^{-1}
  = \alpha + \beta \bm{n},
\end{equation}
which is a spinor of the $\bm{a}$-$\bm{b}$ plane, as desired.
Conversely, we can use the inverse rotation to transform such a spinor
back to the $\bm{x}$-$\bm{y}$ plane.

This well defined method of rotating the spinors explains another
feature of our approach.  To construct the SWSHs, we made a completely
arbitrary choice of basis---in particular of the $\bm{z}$ vector that
appears throughout Sec.~\ref{sec:defining-wigners-d}.  But the
observed field we obtain at the end of this process (\eg, the
gravitational-wave field $h$) is invariant under different choices of
that basis.  In particular, if we have a second basis
$(\bm{x}', \bm{y}', \bm{z}')$, there exists some rotation $\mathbf{\MakeUppercase{R}}_{\text{f}}$ such that
\begin{subequations}
  \label{eq:prime-frame}
  \begin{align}
    \bm{x}' &= \mathbf{\MakeUppercase{R}}_{\text{f}}\, \bm{x}\, \mathbf{\MakeUppercase{R}}_{\text{f}}^{-1}, \\
    \bm{y}' &= \mathbf{\MakeUppercase{R}}_{\text{f}}\, \bm{y}\, \mathbf{\MakeUppercase{R}}_{\text{f}}^{-1}, \\
    \bm{z}' &= \mathbf{\MakeUppercase{R}}_{\text{f}}\, \bm{z}\, \mathbf{\MakeUppercase{R}}_{\text{f}}^{-1}.
  \end{align}
\end{subequations}
We can then define $\mathbf{\MakeUppercase{R}}' \coloneqq \mathbf{\MakeUppercase{R}}\,
\mathbf{\MakeUppercase{R}}_{\text{f}}^{-1}$, so that
\begin{subequations}
  \label{eq:tangent-frame-prime}
  \begin{align}
    \bm{a} &\equiv \mathbf{\MakeUppercase{R}}'\, \bm{x}'\,
    \mathbf{\MakeUppercase{R}}'^{-1}, \\
    \bm{b} &\equiv \mathbf{\MakeUppercase{R}}'\, \bm{y}'\,
    \mathbf{\MakeUppercase{R}}'^{-1}, \\
    \bm{n} &\equiv \mathbf{\MakeUppercase{R}}'\, \bm{z}'\,
    \mathbf{\MakeUppercase{R}}'^{-1},
  \end{align}
\end{subequations}
where $\bm{a}$, $\bm{b}$, and $\bm{n}$ are precisely the same geometric
objects found in Eq.~\eqref{eq:tangent-frame}.  We can go through the
entire process, simply replacing the unprimed basis vectors with their
primed equivalent, evaluating the SWSHs and rotating by
$\mathbf{\MakeUppercase{R}}'$ to get an element of the $\bm{a}$-$\bm{b}$ spinor
space, and obtain precisely the same value.  The result is invariant
with respect to the choice of basis.  Most importantly, we now have
simple and well defined methods for computing and rotating SWSHs.

\section{The differential operator \texorpdfstring{$\eth$}{eth} and
  angular-momentum operators}
\label{sec:eth}

In Sec.~\ref{sec:defining-wigners-d}, we went to some lengths to
express Wigner's $\ensuremath{\mathfrak{D}} $ matrices directly in terms of quaternions,
instead of their more traditional presentation in terms of Euler
angles.  The reason for this is that quaternions form the best
presentation of both the rotation and spin groups: they are simpler
and more intuitive to manipulate; they are more closely linked to the
geometry they describe; they are free from singularities; and they are
more efficient to compute with than angles.  However, some of the more
common tasks when analyzing spherical harmonics and related functions
involve the angular-momentum operator.  This is conventionally given
in terms of Euler angles.  It would be deeply unfortunate if we had to
convert our function back to the Euler-angle presentation whenever we
need to apply angular momentum operators.  Instead, we will see that a
simple geometric argument gives rise to the appropriate operators in
terms of quaternions.  We will also see another natural set of
operators, and find that one of these is identical to the important
spin-raising operator $\eth$ defined by Newman and Penrose in their
original description of spin-weighted spherical
harmonics.\cite{Newman1966}

The familiar idea behind the angular-momentum operator is to find the
rate of change in the value of a function defined on a sphere as one
rotates around that sphere.  More generally, for a function of a
rotation operator, $f(\mathbf{\MakeUppercase{R}})$, we wish to find the rate of
change in the function as we apply infinitesimal rotations to
$\mathbf{\MakeUppercase{R}}$.  We first construct some rotation
$\e^{\theta\, \bm{e}_{j}/2}$ where $\bm{e}_{j}$ is one of the
standard unit basis vectors.  We will apply this rotation to
$\mathbf{\MakeUppercase{R}}$ and differentiate the function with respect to
$\theta$.  That is, we define
\begin{equation}
  \label{eq:ang-mom-ops}
  L_{j}\, f(\mathbf{\MakeUppercase{R}})
  \coloneqq \left. -\bm{z} \frac{\partial}{\partial \theta} f
    \left( \e^{\theta\, \bm{e}_{j}/2}\, \mathbf{\MakeUppercase{R}} \right)
  \right|_{\theta=0}.
\end{equation}
This coincides with our intuitive notion of the angular momentum
operator evaluating the change in an infinitesimal rotation about a
particular axis.  We will show below that this is precisely the
angular-momentum operator as found in the theory of the symmetric
top.\cite{Goldberg1967, Morrison1987} This is a slight generalization
of the more familiar angular-momentum operator, which is usually seen
as a differential operator acting on functions defined on the sphere
$S^{2}$; while this more general operator is defined for functions on
$S^{3}$, it also reduces to the simpler one.  The particular form
given here is even more unusual, however, in that it is not defined in
terms of Euler angles, but directly in terms of elements of the spin
group.

From a purely algebraic perspective the choice to apply the perturbing
rotation by multiplying on the left seems fairly arbitrary.  Arguments
from geometric algebra\cite{Doran2010} suggest that the
\emph{geometric} interpretation for this algebraic operation of
multiplication on the left is to model a physical rotation of the
system; multiplication on the right corresponds to rotation of the
basis with respect to which the function is defined.  For most
physical applications, we are more interested in the former, which is
why $L_{j}$ as defined above is familiar to physicists.  Of course, we
have seen that spin-weighted functions depend explicitly on the basis,
as shown in Eq.~\eqref{eq:spin-weight}.  So a variant form of the
angular-momentum operator will also be useful for our purposes.  We
now define a comparable operator involving multiplication on the
right:
\begin{equation}
  \label{eq:ang-mom-ops-K}
  K_{j}\, f(\mathbf{\MakeUppercase{R}})
  \coloneqq \left. -\bm{z} \frac{\partial}{\partial \theta} f
    \left( \mathbf{\MakeUppercase{R}}\, \e^{\theta\, \bm{e}_{j}/2} \right)
  \right|_{\theta=0}.
\end{equation}
Note that the only difference between $L_{j}$ and $K_{j}$ is the order
of multiplication inside the function argument.  This operator
measures the dependence of the function on the frame in which it was
defined, which is not typically a useful notion in physics, so $K_{j}$
is very uncommon in physics.  Nonetheless, it is useful in the
analysis of spin-weighted functions; we will see that it is (up to an
overall sign) precisely the operator introduced by Goldberg
\etal.\cite{Goldberg1967}

To demonstrate equality between the $L_{j}$ operator given here and
the more common operator given in terms of Euler angles---or between
$K_{j}$ and the one given in terms of Euler angles by Goldberg
\etal---we could express the quaternion argument $\mathbf{\MakeUppercase{R}}$ as a
function of Euler angles, and compare the action of the operators and
show that they must be equal for an arbitrary function.  However,
Euler angles are to be eschewed in all cases.  A more relevant
approach uses the fact that, as mentioned above, the functions
$\ensuremath{\mathfrak{D}} ^{(\ell)}_{m', m}$ form a complete orthogonal system in the space
of square-integrable complex-valued functions on $S^{3}$.  Thus, it is
sufficient to demonstrate equality of the operators when acting on
these functions.  As usual, we define the raising and lowering
operators as $L_{\pm} = L_{x} \pm \bm{z}\, L_{y}$ and
$K_{\pm} = K_{x} \mp \bm{z}\, K_{y}$.  Straightforward evaluation with
Eq.~\eqref{eq:WignerD-naive} shows that
\begin{subequations}
  \label{eq:ang-mom-ops-L-action}
  \begin{align}
    L_{z}\, \ensuremath{\mathfrak{D}} ^{(\ell)}_{m', m}(\mathbf{\MakeUppercase{R}})
    &= m'\, \ensuremath{\mathfrak{D}} ^{(\ell)}_{m', m}(\mathbf{\MakeUppercase{R}}), \\
    L_{\pm}\, \ensuremath{\mathfrak{D}} ^{(\ell)}_{m', m}(\mathbf{\MakeUppercase{R}})
    &= \sqrt{(\ell \mp m') (\ell \pm m' +1)}\,
      \ensuremath{\mathfrak{D}} ^{(\ell)}_{m' \pm 1, m}(\mathbf{\MakeUppercase{R}}),
  \end{align}
\end{subequations}
and similarly
\begin{subequations}
  \label{eq:ang-mom-ops-K-action}
  \begin{align}
    K_{z}\, \ensuremath{\mathfrak{D}} ^{(\ell)}_{m', m}(\mathbf{\MakeUppercase{R}})
    &= m\, \ensuremath{\mathfrak{D}} ^{(\ell)}_{m', m}(\mathbf{\MakeUppercase{R}}), \\
    K_{\pm}\, \ensuremath{\mathfrak{D}} ^{(\ell)}_{m', m}(\mathbf{\MakeUppercase{R}})
    &= \sqrt{(\ell \mp m) (\ell \pm m +1)}\,
      \ensuremath{\mathfrak{D}} ^{(\ell)}_{m', m \pm 1}(\mathbf{\MakeUppercase{R}}).
  \end{align}
\end{subequations}
These are---up to minor sign differences---the same expressions found
elsewhere in the literature,\cite{Goldberg1967, Morrison1987} showing
that our geometric definitions of $L_{j}$ and $K_{j}$ are equivalent
to previous definitions.\footnote{Note that
  Goldberg~\textit{et~al.}\cite{Goldberg1967} defined $K_{j}$ with a
  relative negative sign, and defined the corresponding spin raising
  and lowering operators with another relative sign and conjugation.
  That is, to reproduce their results, we would need to perform the
  translations $K_{z} \leftrightarrow -K_{z}$ and
  $K_{\pm} \leftrightarrow K_{\mp}$.  Their choices appear to have
  been motivated by the curious negative sign in
  $\ensuremath{\scripts{_{s}}{Y}{_{\ell,m}}} \propto \ensuremath{\mathfrak{D}} ^{(\ell)}_{m, -s}$.  The choices here are
  made to enforce the symmetry between Eqs.~\eqref{eq:ang-mom-ops}
  and~\eqref{eq:ang-mom-ops-K}, which leads to symmetry between
  Eqs.~\eqref{eq:ang-mom-ops-L-action}
  and~\eqref{eq:ang-mom-ops-K-action}.}  With the definitions given
here, there is no extraneous conversion to Euler coordinates on the
spin group, for example.

Now, recalling that Eq.~\eqref{eq:SWSH-definition} defined the SWSHs
in terms of $\ensuremath{\mathfrak{D}} ^{(\ell)}_{m', m}$, and that the spin $s$ of a SWSH
corresponds to $-m$, we see that $K_{+}$ is an index-raising operator,
but a spin-\emph{lowering} operator.  Similarly, $K_{-}$ is an
index-lowering operator, but a spin-\emph{raising} operator.  These
are important quantities in the literature on SWSHs, having been
introduced by Newman and Penrose\cite{Newman1966} as $\bar{\eth}$ and
$\eth$, respectively.  We now extend to SWSHs defined on the spin
group:
\begin{subequations}
  \begin{align}
    \eth \left[ \scripts{_{s}}{Y}{_{\ell,m}} (\mathbf{\MakeUppercase{R}}) \right]
    &= -K_{-} \left[\scripts{_{s}}{Y}{_{\ell,m}}(\mathbf{\MakeUppercase{R}})\right],
    \\
    \bar{\eth}\left[\scripts{_{s}}{Y}{_{\ell,m}} (\mathbf{\MakeUppercase{R}})\right]
    &= K_{+} \left[\scripts{_{s}}{Y}{_{\ell,m}} (\mathbf{\MakeUppercase{R}})\right].
  \end{align}
\end{subequations}
These equations are true if we simply copy the original relations for
$\eth$ and $\bar{\eth}$ acting on SWSHs from Newman and Penrose.
However, because the factor of $(-1)^{s}$ in
Eq.~\eqref{eq:SWSH-definition} differs from their convention, it might
also be reasonable to incorporate an additional sign change.

Finally, we can define spin-weighted spherical functions generally.
Acting on an arbitrary SWSH, we see that
\begin{equation}
  \label{eq:Kz-SWSH}
  K_{z}\, \scripts{_{s}}{Y}{_{\ell,m}} (\mathbf{\MakeUppercase{R}})
  = -s\, \scripts{_{s}}{Y}{_{\ell,m}} (\mathbf{\MakeUppercase{R}}).
\end{equation}
That is, $\scripts{_{s}}{Y}{_{\ell,m}}$ is an eigenfunction of
$K_{z}$, with eigenvalue $-s$.  Penrose and Rindler\cite{Penrose1987}
used a comparable relation involving the commutator
$[\eth, \bar{\eth}]$ to define SWSHs, so we follow their example.  We
define a spin-weighted spherical function of weight $s$ to be a
function $\prefixscripts{_{s}}{f}$ taking arguments from the
three-dimensional spin group and mapping to an associated
two-dimensional Euclidean spinor space (which is isomorphic to
$\mathbb{C}$) satisfying
\begin{equation}
  \label{eq:define-SWSF}
  K_{z}\, \prefixscripts{_{s}}{f} = -s\, \prefixscripts{_{s}}{f}.
\end{equation}
The choice of a particular two-dimensional subspace selects a unique
direction $\bm{z}$ orthogonal to it, which is enough to define the
operator $K_{z}$, making this definition independent of any particular
frame chosen to express the functions concretely.

\section{Conclusions}
\label{sec:Conclusions}

This paper has shown that spin-weighted spherical functions (SWSFs)
\emph{cannot} be defined as functions on the sphere $S^{2}$.
Section~\ref{sec:spin-weight-funct} established that the missing
structure is a choice of fiducial direction in the tangent space to
the sphere at each point.  But topological restrictions complicate
such choices, as shown in Sec.~\ref{sec:tangent-frame}.  It turns out
that the Hopf bundle provides the perfect structure for resolving
these complications, simultaneously treating both the sphere $S^{2}$
and the alignment of its tangent spaces.  The Hopf bundle is defined
on the sphere $S^{3}$ (that is its ``entire'' space).  This is also
the topology of the spin group $\mathrm{Spin}(3)$, but the group has
additional structure allowing us to multiply elements and apply other
manipulations.  In particular, the quaternions were used to discuss
the Hopf bundle more easily than the usual Cartesian presentation.

There is some subtlety regarding the use of the spin group rather than
the rotation group $\mathrm{SO}(3)$, but it is clear that either will suffice
for the important cases of integer spin, while the spin group is
needed for half-integer spin.  Thus, the group of unit quaternions was
proposed as the appropriate domain on which to define SWSFs.
Furthermore, the codomain into which SWSFs should map was shown to be
equivalently regarded as the complex numbers $\mathbb{C}$ or as the
spinor space of the $\bm{x}$-$\bm{y}$ plane.  The latter is
more useful, because we can use a geometric construction to create the
standard basis for SWSFs, known as spin-weighted spherical harmonics
(SWSHs).  These are closely related to Wigner's $\ensuremath{\mathfrak{D}} $ matrices, which
we derived purely in terms of quaternions in
Sec.~\ref{sec:defining-wigners-d}.  Then, we saw that the result can
be uniquely rotated into the spinor space tangent to the sphere at a
given point, and thus corresponds to a field propagating orthogonally
to the sphere.

Having shown that SWSFs and SWSHs can be defined and described
entirely within the intuitive and powerful framework of quaternions,
Sec.~\ref{sec:eth} then showed that angular momentum operators can be
defined readily within the same framework.  In particular, the
standard $\bm{L}$ operator is given by applying a rotation to the
argument of a SWSF, and differentiating with respect to the size of
that rotation.  But this rotation is applied on the left; if the
rotation is applied on the right, a distinct operator is found.  It
turns out that this operator $\bm{K}$ is identical (up to minor
quirks of conventions) to an operator introduced by Goldberg
\etal.\cite{Goldberg1967}  As they showed, the ladder operators
associated with $\bm{K}$ are essentially the same as the
spin-raising and -lowering operators $\eth$ and $\bar{\eth}$ defined
by Newman and Penrose in their original introduction of
SWSHs.\cite{Newman1966}

Defining SWSFs as functions on the spin group first serves the basic
function of actually providing a mathematically well posed formulation
of SWSFs---an objective that has been surprisingly rare in the
literature.  But on a more practical level, it allows us to transform
SWSFs.  By using quaternions, we further provide a unified system that
combines algebraic power, computational benefits, and geometric
interpretations.  For example, quaternions are a special case of the
spacetime algebra\cite{Hestenes2003a} which allows us to treat boosts
and rotations in the same language.  This approach was previously used
to transform SWSHs under boosts\cite{Boyle2015} by projecting the
spacetime algebra down to quaternion components, and evaluating the
SWSHs as functions of those quaternions.  This is one example of the
power and simplifications that result from defining SWSFs as functions
of quaternions.

A related construction is that of tensorial spin-weighted spherical
harmonics, which arise in various calculations for general
relativity.\cite{Newman2006} Essentially, these harmonics are produced
by coupling SWSFs to symmetric trace-free tensor fields; in
particular, a basis is given by coupling the SWSHs to various tensor
products involving the vector fields $\bm{n}$, $\bm{m}$, and
$\bar{\bm{m}}$ defined above.  As we've seen there are natural
maps taking the spin group into these vector fields.  Alternatively,
simple tensor products can be constructed of multiple copies of
$\bm{m}$, and the spin-weight of each such product lowered to give
rise to the full variety of tensorial spin-weighted harmonics.  But
this requires applying $\bar{\eth}$ to the tensor product, which in
turn requires it to be defined on vector fields.  Our definitions of
$\bar{\eth}$ relied on the angular-momentum operator $K_{j}$ given in
Eq.~\eqref{eq:ang-mom-ops-K}, which is only defined on spinor fields.
Fortunately, we can simply regard the vector fields as their
equivalent quaternion fields, and the same formula applies, resulting
in expressions identical to those found by Newman and
Silva-Ortigoza.\cite{Newman2006} Thus, the tensorial SWSHs could also
be coherently defined as functions from the spin group.

The final few paragraphs of Sec.~\ref{sec:eth} (among others)
illustrate that the various signs and other conventions lead into a
confusing morass of inconsistent definitions when trying to compare
results from different papers.  In practice, the only reliable way to
determine these conventions is to derive everything directly.  The
holistic approach of defining spin-weighted functions as functions on
the spin group eases this difficulty by unifying the algebra with the
geometry it represents.

\begin{acknowledgments}
  It is my pleasure to thank Marc Kamionkowski, Larry Kidder, and Saul
  Teukolsky for helpful conversations, and Barry Wardell for his help
  sorting out various conventions involved in Wigner's $\ensuremath{\mathfrak{D}} $ matrices.
  I particularly thank Leo Stein for a very helpful and thorough
  critique of an earlier draft of this paper.  This project was
  supported in part by the Sherman Fairchild Foundation, and by NSF
  Grants No.\ PHY-1306125 and AST-1333129.
\end{acknowledgments}

\appendix 

\section{Evaluating polynomials with factorial and binomial
  coefficients}
\label{sec:eval-polyn-with}

It is well known that polynomials should be evaluated numerically in
Horner form, which is both faster and more accurate than naive direct
evaluation.\cite{Knuth1997, Press2007}  In this appendix, a few
specializations of Horner form are presented, which can be used when
the coefficients of the polynomial are factorials, or products and
ratios of factorials---including binomials.  These specializations
work by calculating the coefficients themselves as part of the
algorithm.  In addition to the standard benefits of Horner-form
evaluation, this speeds the computation of the coefficients and avoids
arithmetic overflow when very large factorials are involved.

We first review the standard Horner form.  Given a polynomial with
coefficients $c_{j}$, the polynomial may be evaluated as
\begin{equation}
  \label{eq:Horner-form}
  \sum_{j=j_{1}}^{j_{2}} c_{j} x^{j} = x^{j_{1}}
  \left\{c_{j_{1}} + x \left[ c_{j_{1}+1} + x
      \left(c_{j_{1}+2} + \ldots \right) \right] \right\}.
\end{equation}
Explicitly evaluating the left-hand side involves redundant
calculations for the powers of $x$, since $x^{j}$ requires knowledge
of $x^{j-1}$.  The summation itself can also lead to delicate
cancellations between terms, which can destroy numerical accuracy.  If
instead we evaluate the right-hand side, there is no redundant
calculation of powers of $x$, and numerical accuracy is retained.
This simple algorithm is given as follows.
\begin{algorithmic}[1]
  \State $p = c_{j_{2}}$
  \State $j = j_{2}-1$
  \While{$j \geq j_{1}$}
    \State $p = c_{j} + x \, p$
    \State $j = j-1$
  \EndWhile
  \State $p = x^{j_{1}} \, p$
\end{algorithmic}
The final value of $p$ is the value of the polynomial.  In
this case, we need some way to obtain the values of the coefficients,
which may be done by either indexing an array of precomputed
coefficients or by using a function that computes the coefficient
given the index.

Now, if the coefficient is a factorial in the summation index $j$,
there is another redundancy, much like the redundancy in computing
various powers of $x^{j}$.  That is, computing $j!$ requires knowing
$(j-1)!$, and so on.  We can take advantage of this structure to find
a more efficient expression.  Generalizing slightly, we assume there
is still some coefficient, but we have factored out a factorial:
\begin{equation}
  \label{eq:Horner-form-factorial}
  \sum_{j=j_{1}}^{j_{2}} c_{j}\, j!\, x^{j} = j_{1}!\, x^{j_{1}}
  \left\{c_{j_{1}} + (j_{1}+1) x \left[ c_{j_{1}+1} + \ldots \right]
  \right\}.
\end{equation}
Here, the algorithm is just slightly different; line~4 above becomes
\begin{algorithmic}[1]
  \setalglineno{4}
  \State \hspace{\algorithmicindent} $p = c_{j} + (j+1) \,
  x \, p$
\end{algorithmic}
and the final result in this case only will also need to be multiplied
by $j_{1}!$.

It is a trivial task to modify this algorithm to compute the sum
$\sum_{j} c_{j} (A+j)! x^{j}$.  For our purposes, a more interesting
sum is
\begin{equation}
  \label{eq:Horner-form-factorial-inverse}
  \sum_{j=j_{1}}^{j_{2}} c_{j} \frac{(A+j_{1})!} {(A+j)!}\, x^{j} =
  x^{j_{1}}  \left\{c_{j_{1}} + \frac{1} {A+j_{1}+1} x \left[
      c_{j_{1}+1} + \ldots \right] \right\},
\end{equation}
for which line~4 in the algorithm becomes
\begin{algorithmic}[1]
  \setalglineno{4}
  \State \hspace{\algorithmicindent}
  $p = c_{j} + \frac{1} {A+j+1} \, x \, p$
\end{algorithmic}
An interesting modification occurs when the sign of $j$ is flipped in
the factorial:
\begin{equation}
  \label{eq:Horner-form-factorial-inverse-negative}
  \sum_{j=j_{1}}^{j_{2}} c_{j} \frac{(B-j_{1})!} {(B-j)!}\, x^{j} =
  x^{j_{1}} \left\{c_{j_{1}} + (B-j_{1}) x \left[ c_{j_{1}+1} +
      (B-j_{1} - 1) x \left( c_{j_{1}+2} + \ldots \right) \right]
  \right\},
\end{equation}
in which case that line in the algorithm should be
\begin{algorithmic}[1]
  \setalglineno{4}
  \State \hspace{\algorithmicindent}
  $p = c_{j} + (B-j) \, x \, p$
\end{algorithmic}
Note that we \emph{multiply} by $B-j$, whereas we were \emph{dividing}
by $A+j+1$ in the previous case.

The most important generalization of this algorithm applies to
multiple factorials.  For example, we can compute a polynomial with
binomial coefficient as
\begin{equation}
  \label{eq:binomial-horner}
  \sum_{j} c_{j}\, \binom{C}{j}\, x^{j} = \binom{C}{j_{1}}\, \sum_{j}
  c_{j}\, \frac{j_{1}!\, (C-j_{1})!} {j!\, (C-j)!}\, x^{j}
\end{equation}
using the same algorithm with
\begin{algorithmic}[1]
  \setalglineno{4}
  \State \hspace{\algorithmicindent}
  $p = c_{j} + \frac{C-j} {j+1} \, x \, p$
\end{algorithmic}
and, of course, multiplying the final $p$ by $\binom{C}{j_{1}}$.

For the purposes of this paper, it is interesting to apply this to
computing elements of Wigner's $\ensuremath{\mathfrak{D}} $ matrices.  The sum in question
[depending on which branch is taken in Eq.~\eqref{eq:WignerD-general}]
looks something like this:
\begin{widetext}
  \begin{align}
    \sum_{j = j_{1}} ^{j_{2}} \binom{\ell+m'} {\ell-m-j}\,
    \binom{\ell-m'} {j}\, x^{j}
    &=
      \binom{\ell+m'} {\ell-m-j_{1}}\, \binom{\ell-m'} {j_{1}}\,
      \sum_{j = j_{1}} ^{j_{2}} \frac{(\ell-m-j_{1})!} {(\ell-m-j)!}
      \frac{(m'+m+j_{1})!} {(m'+m+j)!} \frac{j_{1}!} {j!}
      \frac{(\ell-m'-j_{1})!} {(\ell-m'-j)!}\, x^{j}.
  \end{align}
\end{widetext}
Noting that the $c_{j}$ coefficients in this case are all $1$, this
sum can be computed just as easily as the cases above using
\begin{algorithmic}[1]
  \setalglineno{4}
  \State \hspace{\algorithmicindent}
  $p = 1 + \frac{\ell-m-j} {m'+m+j+1} \,
  \frac{\ell-m'-j} {j+1} \, x \, p$
\end{algorithmic}
It is, of course, possible to use Horner's rule to evaluate the
polynomial directly, taking each coefficient as the appropriate
binomial.  However, that requires calculating two binomial
coefficients at each step, which may be reasonably accomplished by
indexing a pre-computed array of values.  That, in turn, requires
additional calculations to compute the index, as well as non-local
operations in memory to retrieve the correct value.

Unfortunately, we can still find values for which the sum is not well
conditioned, leading to a loss in accuracy when implemented either
with Horner's original algorithm or with this specialized algorithm.
For example, if $x \approx -1$, $\ell$ is large, and $m' \approx -m$
is close to $0$, then the coefficient
$\frac{\ell-m-j} {m'+m+j+1} \, \frac{\ell-m'-j} {j+1}$
will vary dramatically as $j$ changes.  At some point during the loop
over $j$, this will bring the value of $p$ very close to $-1$, leading
to rapid loss of accuracy in the sum.  Perhaps even more importantly,
there are simply many terms in the sum when $\ell$ is large.  This is
never a problem for $\ell \lesssim 12$, because only one digit of
precision is lost.  For larger values of $\ell$, however, we might
expect these sums to lose up to $\log_{10}(2^{\ell+1}/\pi\ell)$ digits
of accuracy\cite{Choi1999}---though that appears to overestimate the
error by one or two digits when using this new algorithm.  In any
case, accuracy appears to be completely lost for certain values of
$x$, $m'$, and $m$ when $\ell$ is as low as $\ell \approx 60$.  If
accurate values are needed for large $\ell$, the results of this
algorithm can be used as input to initialize recursion relations,
which retain accuracy for all values of $m$ and $m'$.\cite{Choi1999}


\section{Parametrizing SWSFs}
\label{sec:param-spher}

In practice, when using SWSFs to describe radiation fields, we need a
useful parametrization.  We now discuss various parametrizations that
can be used for this purpose, starting from a fairly general
perspective.  We want some parameter space $P$ along with
\emph{continuous} maps $\mathfrak{e}$ and $\mathfrak{b}$ as
given here:
\begin{equation}
  \label{eq:ParametrizationMaps}
  \raisebox{-0.5\height}{\includegraphics{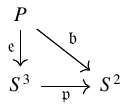}}
\end{equation}
Note that this diagram need not be commutative.  Indeed, in some cases
we will see that $\mathfrak{e}$ may not be defined on all of $P$.
Ideally, we want $\mathfrak{e}$ to be defined throughout $P$ and
injective (distinct elements in $P$ map to distinct elements of
$S^{3}$), so that the spin-weighted function will be single-valued.
We also want $\mathfrak{b}$ to be surjective (for each element of
$S^{2}$, there is some element of $P$ that maps to it), so that the
function can describe observations in each direction.

\begin{turnpage}
\begin{table*}[ht]
  \renewcommand{\arraystretch}{2.2}
  \begin{ruledtabular}
    \begin{tabular}{lccc}
      \rule[-10pt]{0pt}{0pt}%
      Name & Parameter space $P$ & $\mathfrak{e}:P \to S^{3}$
      & $\mathfrak{b}: P \to S^{2}$
      \\[0pt] \hline \rule{0pt}{12pt}%
      Spherical coordinates of $S^{2}$
         & $(\vartheta, \varphi) \in I \times S^{1}$
         & $ \displaystyle
           \exp\left[ \frac{\varphi}{2}\, \bm{z} \right]\,
           \exp\left[\frac{\vartheta}{2}\, \bm{y} \right]$
         & $ \displaystyle
           \Big(\sin\vartheta \cos\varphi, \sin\vartheta \sin\varphi,
           \cos\vartheta \Big)$
      \\
      Stereographic coordinates\footnote{These expressions are correct
        for projection from the south pole; it is more common to see
        projection from the north pole.  This choice is made for
        consistency with our previous definitions, which give the
        points of the sphere as rotations of $\bm{z}$ (the north
        pole).} of $S^{2}$
         & $\zeta \in \mathbb{C} \cup \{\infty\}$
         & $\displaystyle
           \frac{1} {\sqrt{1+\zeta\bar{\zeta}}} \left[ 1
             + \i \frac{\zeta-\bar{\zeta}} {2} \bm{x}
             + \frac{\zeta + \bar{\zeta}} {2} \bm{y}  \right]$
         & $\displaystyle
           \left( \frac{\zeta+\bar{\zeta}} {1+\zeta\bar{\zeta}},
           \frac{\zeta-\bar{\zeta}}
           {\i(1+\zeta\bar{\zeta})}, \frac{1-\zeta\bar{\zeta}}
           {1+\zeta\bar{\zeta}} \right)$
      \\
      Homogeneous stereographic coordinates of $S^{2}$
         & $(a+\i b, c+\i d) \in \mathbb{C}^{2} \setminus\{(0,0)\}$
         & $\displaystyle
           \frac{a + d \bm{x} + c \bm{y} + d \bm{z}}
           {\sqrt{a^{2} + b^{2} + c^{2} + d^{2}}}$
         & $\displaystyle
           \frac{1} {r^{2}} \left(
           2(ac + bd),
           2(ad - bc),
           r^{2} - 2(a^{2} + b^{2})
           \right)$
           \raisebox{2ex}{\footnotemark[2]}
           \footnotetext[2]{This expression uses the quantity
           $r^{2} \coloneqq a^{2} + b^{2} + c^{2} + d^{2}$ for
           compactness.}
      \\
      Stereographic coordinates of $S^{3}$
         & $(a, b, c) \in \mathbb{R}^{3} \cup \{\infty\}$
         & $\displaystyle
           \frac{1} {1+r^{2}} \left[1-r^{2} + 2a \bm{x}
           + 2b \bm{y} + 2c \bm{z} \right]$
           \raisebox{2ex}{\footnotemark[3]}
           \footnotetext[3]{This expression uses the quantity
           $r^{2} \coloneqq a^{2} + b^{2} + c^{2}$ for compactness.}
         & $\displaystyle
           \mathfrak{h} \circ \mathfrak{e}$
      \\
      Quaternion logarithms
         & $(x, y, z) \in \mathbb{R}^{3}$
         & $\displaystyle
           \exp[x\bm{x}+y\bm{y}+z\bm{z}]$
         & $\displaystyle
           \mathfrak{h} \circ \mathfrak{e}$
      \\
      Euler angles
         & $(\alpha, \beta, \gamma) \in S^{1} \times I \times S^{1}$
         & $\displaystyle
           \exp\left[\frac{\alpha}{2} \bm{z} \right]\,
           \exp\left[\frac{\beta}{2} \bm{y} \right]\,
           \exp\left[\frac{\gamma}{2} \bm{z} \right]$
         & $\displaystyle
           \mathfrak{h} \circ \mathfrak{e}$
      \\
      Hyperspherical coordinates of $S^{3}$
         & $(\psi, \theta, \phi) \in S^{1} \times I \times S^{1}$
         & $\displaystyle
           \exp \left[ \frac{\psi}{2}\, \left(
             \bm{x} \sin\theta \cos\phi
           + \bm{y} \sin\theta \sin\phi
           + \bm{z} \cos\theta \right) \right]$
         & $\displaystyle
           \mathfrak{h} \circ \mathfrak{e}$
      \\
      Hopf coordinates of $S^{3}$
         & $(\eta, \zeta, \xi) \in I \times S^{1} \times S^{1}$
         & $\displaystyle
           \sin\eta (\cos\zeta + \bm{x} \sin\zeta)
           + \cos\eta (\bm{y} \cos\xi + \bm{z} \sin\xi)$
         & $\displaystyle
           \mathfrak{h} \circ \mathfrak{e}$
      \\
      Invertible quaternions
         & $(w, x, y, z) \in \mathbb{R}^{4} \setminus \{0\}$
         & $\displaystyle
           \frac{w + x\bm{x} + y\bm{y} + z\bm{z}}
           {\sqrt{w^2+x^2+y^2+z^2}}$
         & $\displaystyle
           \mathfrak{h} \circ \mathfrak{e}$
    \end{tabular}
  \end{ruledtabular}
  \caption{\label{tab:parametrizations}%
    \textbf{Parametrizations of SWSFs}.  We
    list a variety of parametrizations that can be used for SWSFs
    along with their mappings into $S^{3}$, considered to be
    parametrized by unit quaternions, and $S^{2}$ considered as a
    subset of $\mathbb{R}^{3}$.  Note that for each of the
    parametrizations of $S^{3}$, we simply use the Hopf map of
    Eq.~\eqref{eq:HopfMap-quaternions} to go from $S^{3}$ to $S^{2}$;
    this would also be possible for the parametrizations of $S^{2}$,
    but they are more normally defined directly.}
\end{table*}
\end{turnpage}

In Table~\ref{tab:parametrizations}, we collect a variety of common
parametrizations along with their respective mappings.  Perhaps
surprisingly, the standard spherical coordinates provide one of the
most useful parametrizations for SWSHs.  They cover the entire sphere
($\mathfrak{b}$ is surjective).  And despite the well-known
coordinate singularities at the north and south poles, each pair
$(\vartheta, \varphi)$ corresponds to a unique element of the spin
group ($\mathfrak{e}$ is injective).  These are the criteria
mentioned above for a useful parametrization, which is why spherical
coordinates may be chosen for implementations of numerical routines
involving SWSHs---such as the state-of-the-art \texttt{spinsfast}
package.\cite{Huffenberger2010, Huffenberger}

It is somewhat more common in the literature on SWSHs to find
stereographic coordinates used.  These have certain advantages over
spherical coordinates, in that they are complex and so lend themselves
to more elegant algebraic manipulations in many cases.  However,
inspection of the mappings shown in the table shows that
$\mathfrak{b}:P \to S^{2}$ is only surjective when the point at
infinity is included, but $\mathfrak{b}:P \to S^{3}$ is not
defined at this point.  In fact, it can be proven that such a mapping
would again violate the Hairy-Ball Theorem.  Thus, stereographic
coordinates---at least in this simple form---do not provide a useful
parametrization except for theoretical applications that do not
require rigorous coverage of $S^{2}$.

Various related constructions may be more capable for certain
purposes, such as using two complementary coordinate patches.  The
complications of implementing this construction are likely not worth
the effort.  The ``homogeneous'' stereographic coordinate system
solves these problems using a pair of complex numbers, which are not
both zero.\cite{Penrose1987} Only their ratio is used, so this is the
projective space $\mathbb{C}\mathrm{P}^{1}$, which is topologically homeomorphic to
$S^{2}$.  This is the formulation favored by Eastwood and
Tod.\cite{Eastwood1982} It is closely related to the 2-spinor
formalism,\cite{ODonnell2003} which is in turn closely related to the
quaternion formalism.  In fact, the two complex components of the
2-spinor $(\pi_{0}, \pi_{1})$ can be considered precisely the
symmetric and antisymmetric parts of the quaternion $\mathbf{\MakeUppercase{R}}$
under conjugation by $\bm{z}$ [see Eqs.~\eqref{eq:quaternion-parts}].
Then, the point on $S^{2}$ described by the stereographic coordinate
$\zeta = \pi_{0} / \pi_{1}$ is precisely the same point as
$\mathbf{\MakeUppercase{R}}\, \bm{z}\, \mathbf{\MakeUppercase{R}}^{-1}$ (that is, the image of
$\mathbf{\MakeUppercase{R}}$ under the Hopf map).  Numerous other parallels exist
between quaternion algebra and the 2-spinor algebra, which may make it
seem at first sight as though the homogeneous coordinates of $\mathbb{C}\mathrm{P}^{1}$
are somehow equivalent to the quaternion representation.  However, the
projective operation loses information, and requires a choice of
algebraic structure that is not inherently present in the quaternions.

A slight generalization of the usual stereographic coordinates of
$S^{2}$ may be more relevant to our purposes, by representing $S^{3}$
directly.  The stereographic projection generalizes to arbitrary
dimensions, so the special case of $S^{3} \subset \mathbb{R}^{4}$ is
straightforward.  Again, however, the point at infinity must be
included in order for this to be a complete parametrization.  In
addition, the algebraic manipulations possible with complex numbers do
not generalize very immediately to this system---though it is not
entirely clear how much of a disadvantage that may be.  On the other
hand, the formulas involved in this parametrization are slightly
simpler than those for the previous one.

In fact, the stereographic coordinates of $S^{3}$ may be shown to be a
simple rescaling of the more familiar quaternion logarithms.  These
logarithms are the generators of rotation, and can be exponentiated to
give unit quaternions.  This exponentiation is periodic in the
magnitude of the logarithm by $2\pi$, in precisely the same way as the
usual complex logarithm is periodic by $2\pi$.  Essentially, the
logarithm is the compactified version of the stereographic coordinates
of $S^{3}$, under the mapping
\begin{equation}
  \label{eq:stereographic-compactification}
  (a,b,c) \mapsto (x,y,z) = (a,b,c)
  \frac{\arccos \frac{1-r^{2}} {1+r^{2}}} {r},
\end{equation}
where $r = \sqrt{a^{2}+b^{2}+c^{2}}$.  Clearly the quaternionic system
is more computationally useful.  Moreover, we recover algebraic
niceties when using quaternions---potentially helpful for theoretical
purposes.  In fact, the quaternion logarithm is very closely
identified with the axis-angle representation: the logarithm is just a
unit vector along the axis, multiplied by one half the angle of the
rotation.

We also have a set of three coordinate systems defined on
$P = S^{1} \times I \times S^{1}$ (or a permutation of those spaces):
the Hopf, hyperspherical, and Euler coordinates.  In each case, the
coordinates are angles that are combined in different ways to arrive
at a rotation.  For example, the hyperspherical coordinates of $S^{3}$
are closely related to the axis-angle representation, where $\psi$
represents the angle, and $(\theta, \phi)$ the spherical-coordinate
direction of the axis.  Unfortunately, all three systems share the
same basic failing: they all provide non-injective mappings to
$S^{3}$.  In the case of the Euler angles, this well known problem is
referred to as ``gimbal lock''.

Finally, we come to what is clearly the preferred representation of
rotations: quaternions.  Quaternions contain clear geometric meaning
that corresponds readily with the axis-angle understanding of
rotations.  They are free from coordinate singularities, and can
simply be multiplied to express the composition of rotations.  Their
algebraic structure is as close as one may come to simple complex
numbers while being appropriately non-commutative.  The quaternions
can be readily normalized if desired, but this is unnecessary as long
as the rotation is applied using the inverse as in
Eq.~\eqref{eq:rotor-action}.  They map surjectively onto both $S^{3}$
and $S^{2}$, and the mapping onto $S^{3}$ is injective modulo
normalization.  These are some of the reasons quaternions were chosen
as the language of rotations for this paper.




\vfil


\let\c\Originalcdefinition
\let\d\Originalddefinition
\let\e\Originaledefinition
\let\i\Originalidefinition

\bibliography{ZotReferences,References}


\end{document}